\documentclass[aps,showpacs,twocolumn,superscriptaddress]{revtex4}
\usepackage{color}
\usepackage{amsfonts}
\usepackage{amssymb}
\usepackage{amsmath}
\usepackage{epsfig}
\usepackage{graphicx}
\usepackage{enumerate}
\usepackage{multirow}
\usepackage{verbatim}
\usepackage{color}
\usepackage{subfigure}

\begin{document}

\title{Topological flat band models with arbitrary Chern numbers}

\author{Shuo Yang}
\affiliation{
Condensed Matter Theory Center and Joint Quantum Institute,
Department of Physics, University of Maryland, College Park, MD 20742, USA
}
\author{Zheng-Cheng Gu}
\affiliation{
Kavli Institute for Theoretical Physics, University of California, Santa Barbara, CA 93106, USA
}
\author{Kai Sun}
\affiliation{
Condensed Matter Theory Center and Joint Quantum Institute,
Department of Physics, University of Maryland, College Park, MD 20742, USA
}
\author{S. Das Sarma}
\affiliation{
Condensed Matter Theory Center and Joint Quantum Institute,
Department of Physics, University of Maryland, College Park, MD 20742, USA
}

\begin{abstract}
We report the theoretical discovery of a systematic scheme to produce topological flat bands (TFBs) with arbitrary Chern numbers. We find that generically a multi-orbital high Chern number TFB model can be constructed by considering multi-layer Chern number $C=1$ TFB models with enhanced translational symmetry. A series of models are presented as examples, including a two-band model on a triangular lattice with a Chern number $C=3$ and an $N$-band square lattice model with $C=N$ for an arbitrary integer $N$. In all these models, the flatness ratio for the TFBs is larger than $30$ and increases with increasing Chern number. In the presence of appropriate inter-particle interactions, these models are likely to lead to the formation of novel Abelian and Non-Abelian fractional Chern insulators. As a simple example, we test the $C=2$ model with hardcore bosons at $1/3$ filling 
and 	an intriguing fractional quantum Hall state is observed.
\end{abstract}


\maketitle

{\it Introduction.} The experimental fractional quantum Hall effect (FQHE) arises from the highly degenerate Landau levels of continuum 2D electron systems, and is described by variational wave functions, first proposed by Laughlin for the primary FQHE states~\cite{Laughlin1983} and later generalized by Jain for composite fermion states~\cite{Jain1989}, which are analytic functions of the 2D spatial coordinates. Many important properties of the FQHE, for example, the hierarchy structures and fractionalized excitations~\cite{Haldane1983,Halperin1984}, can be understood within this framework. It even leads to the predictions of intriguing non-Abelian fractional quantum Hall (FQH) states at certain filling fractions~\cite{HaldaneFQH1988,Moore1991,Read1999,Bonderson2008}. Moreover, based on a classification of the pattern of zeros of symmetric (analytic) polynomials, a systematic way to classify FQH states~\cite{Wenpattern1,Wenpattern2} has been proposed. Thus, our current theoretical knowledge of FQHE is based on the analytic structure the 2D Landau level (LL) Hilbert space.

A key question of fundamental importance in this context is whether the analytic structure of the candidate microscopic wavefunction~\cite{Laughlin1983} associated with the LL subspace is essential to the existence of the FQHE phenomenon. The answer turns out to be no. Conceptually, we know that the essential physics of the FQHE can be regarded as the emergence of a macroscopic topological order~\cite{Wen1990} at low energy, which is, in principle, independent of the microscopic details of the underlying wavefunction. Very recently, it has been shown that various Abelian and non-Abelian FQH states (also known as fractional Chern insulators) can be realized in a large class of so-called topological flat band (TFB) lattice models~\cite{Tang2010,Neupert2010,sun} without any Landau levels~\cite{Sheng2011,Wang2011,Bernevig2011,BernevigNA2011,Wang2012,Yang2012,McGreevy2012,Wu2011}. The basic idea of the TFB models is the following: (a) Use a 2D lattice-based tight-binding topological-band model to mimic the nontrivial topology of a LL characterized by a nonzero Chern number~\cite{Thouless1982}, as first proposed by Haldane~\cite{Haldane1988}, and (b) find a parameter region to realize a narrow bandwidth with a smooth Berry curvature to quench the kinetic energy. 
These new theoretical flatband FQHE discoveries not only improve our understanding about the nature of the FQHE, but also provide us new ways to realize the FQHE in solid state materials~\cite{Venderbos2011,Hu2011,Xiao2011,Venderbos2012,Ghaemi2011} and ultracold atomic systems~\cite{sun,Wu2007}. 

Although considerable progress has been made along this new direction, the observed FQH states in TFB systems have so far been more or less expected since all of them can be realized in a LL. This is mainly because the Chern number of all these TFB models is $C=1$. 
To explore new physics beyond single LLs, it is natural to consider TFB models with higher Chern numbers~\cite{Fa2011,Maissam2011}. However, higher Chern number TFBs with large flatness ratio (band gap/bandwidth) are much harder to obtain without long range hopping. Very recently, a TFB with Chern number $C=2$ (with a flatness ratio $15$) was proposed~\cite{WangHC2012}.

In this Rapid Communication, we propose a generic and systematic scheme to produce arbitrary Chern number TFBs with short-range hopping using multi-orbital structures.
We believe that these high Chern number TFBs will produce novel topological phases when proper interactions are introduced.
As a simple example, we test the $C=2$ model with hardcore bosons at $1/3$ filling and
an intriguing FQH state is observed. Moreover, TFBs with Chern number $C=2$ have the potential to realize many new bilayer non-Abelian FQH states~\cite{Maissam2010} predicted by the pattern of zeros classification and TFBs with Chern numbers $C=8n$ ($n$ is an integer) may result in a class of bosonic integer quantum Hall (IQH) states (known as Kitaev's $E_8$ states~\cite{Kitaev}).

\begin{figure}[tbp]
\begin{center}
\vskip 0.0cm \hspace*{-0.0cm}
\subfigure[]{\includegraphics[width=0.45\columnwidth]{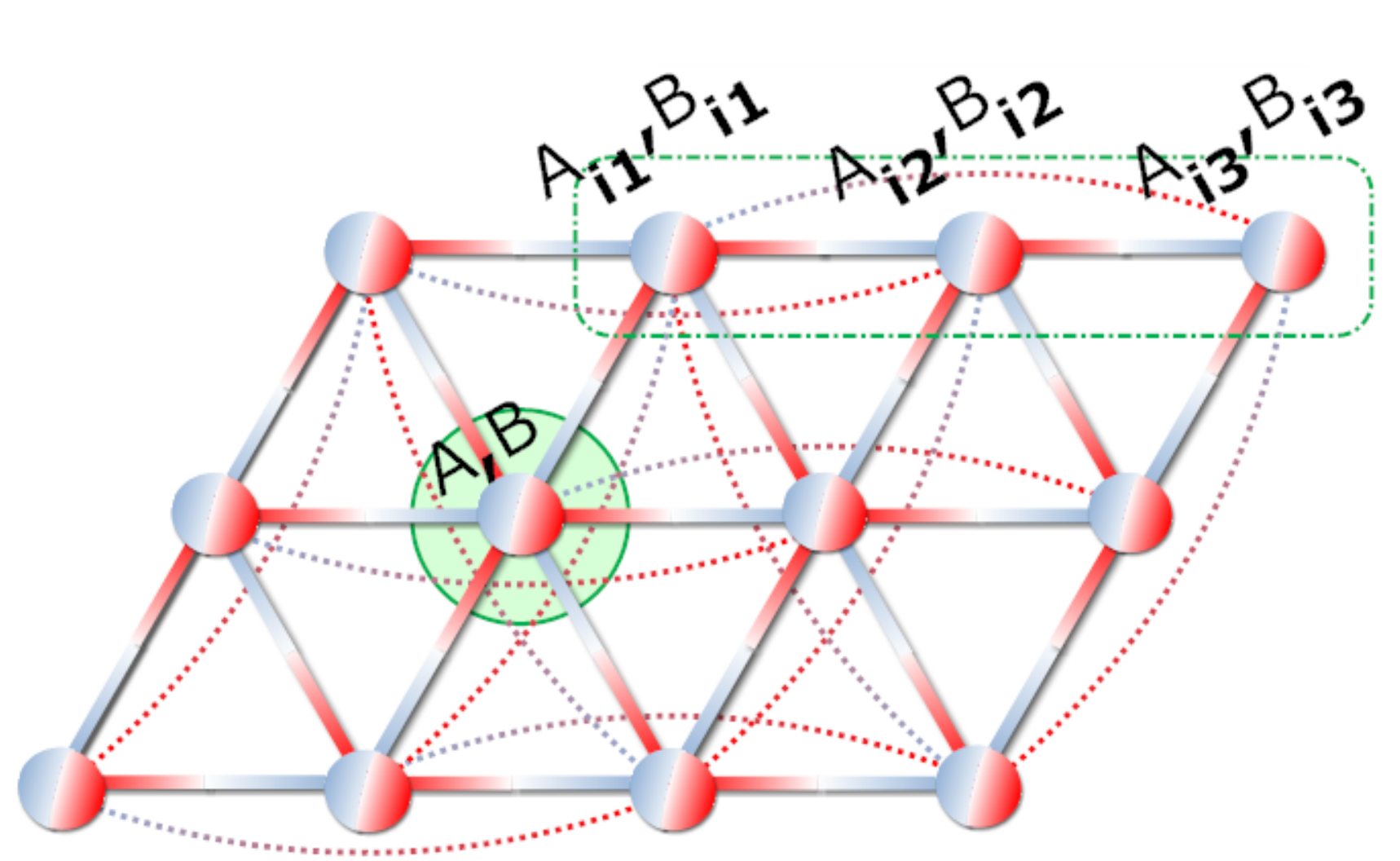}\label{fig:modelC3_NN}}
\subfigure[]{\includegraphics[width=0.45\columnwidth]{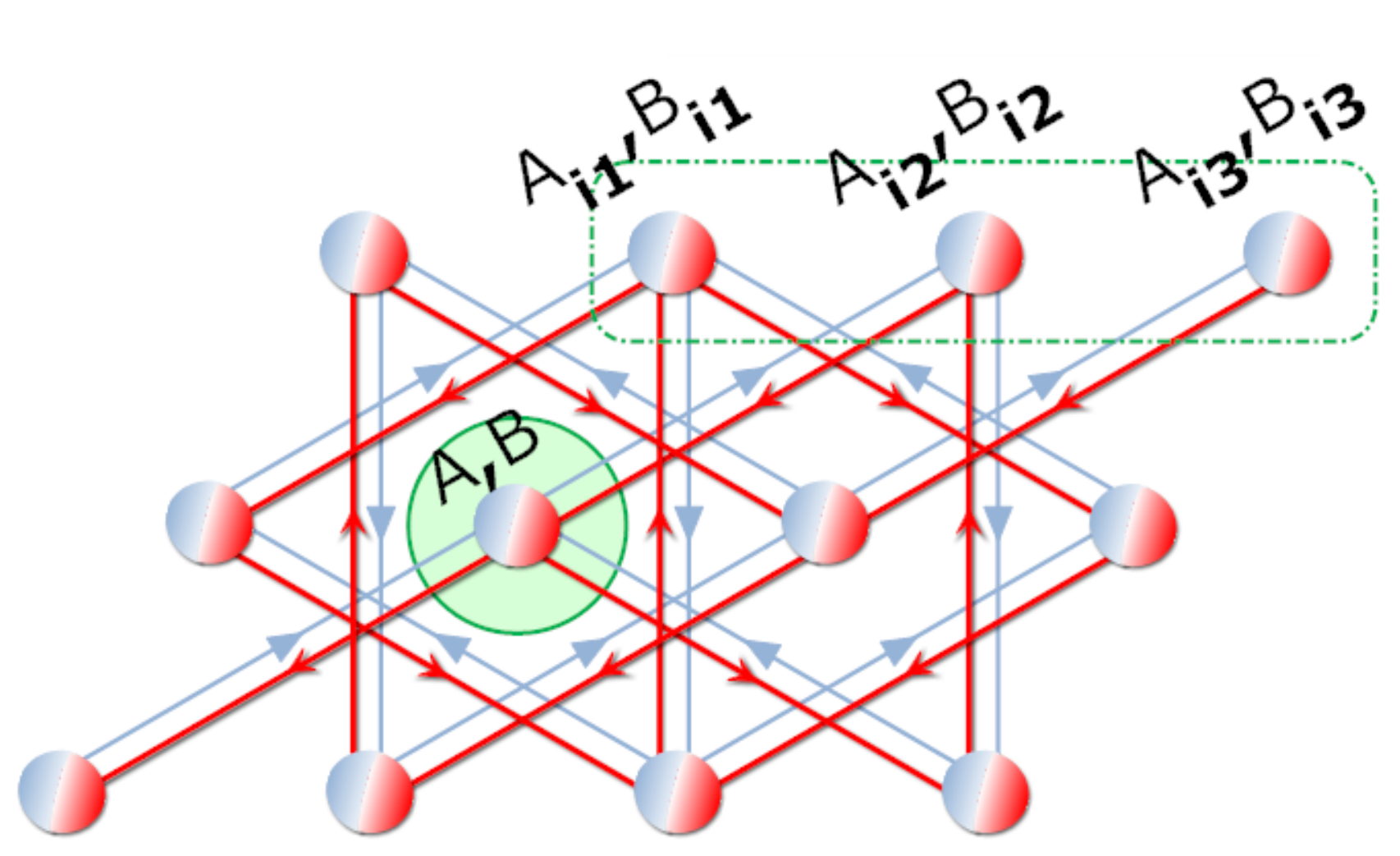}\label{fig:modelC3_NNN}}
\subfigure[]{\includegraphics[width=0.4\columnwidth]{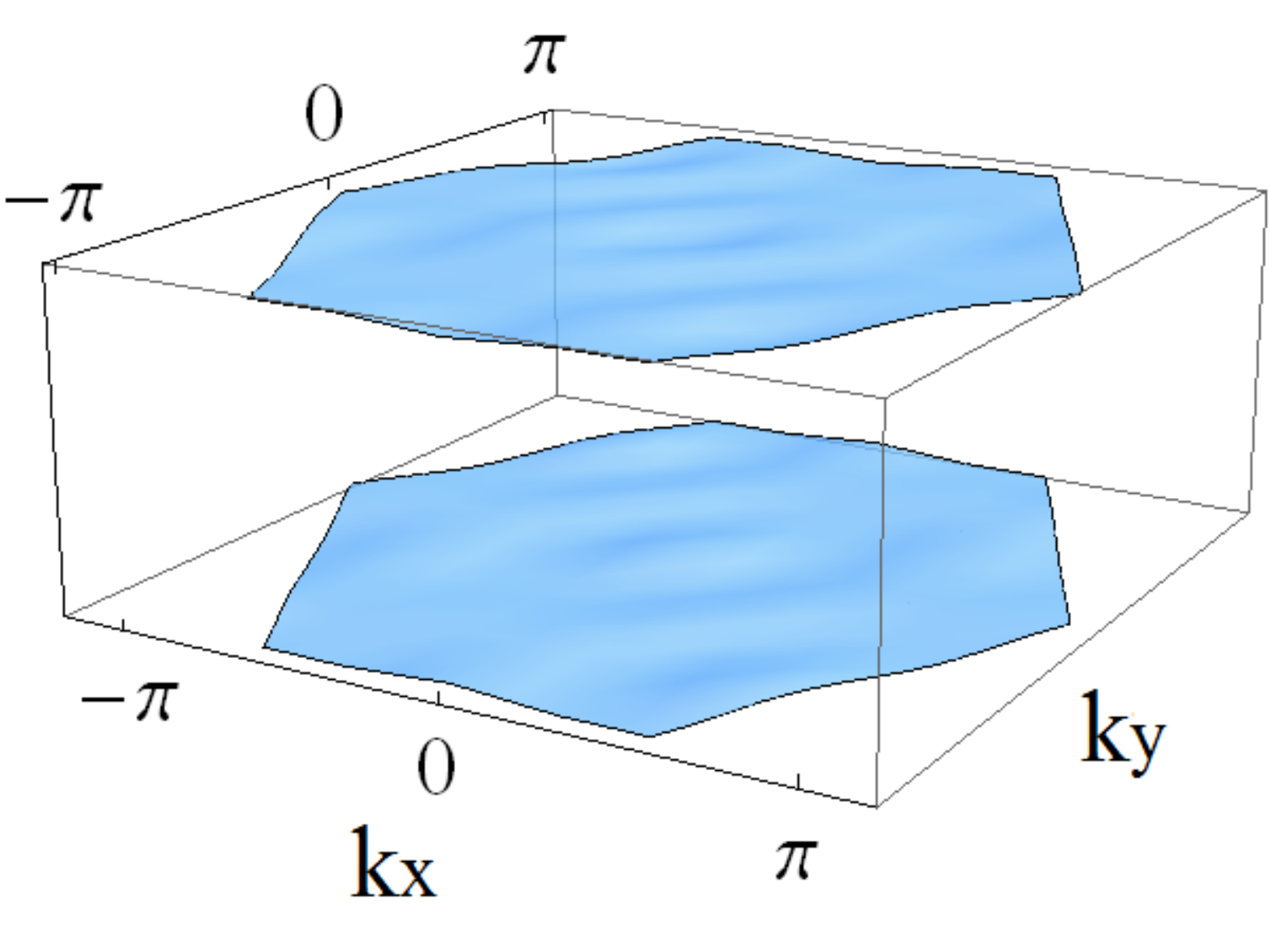}\label{fig:modelC3_band}}
\hspace*{0.3cm}
\subfigure[]{\includegraphics[width=0.4\columnwidth]{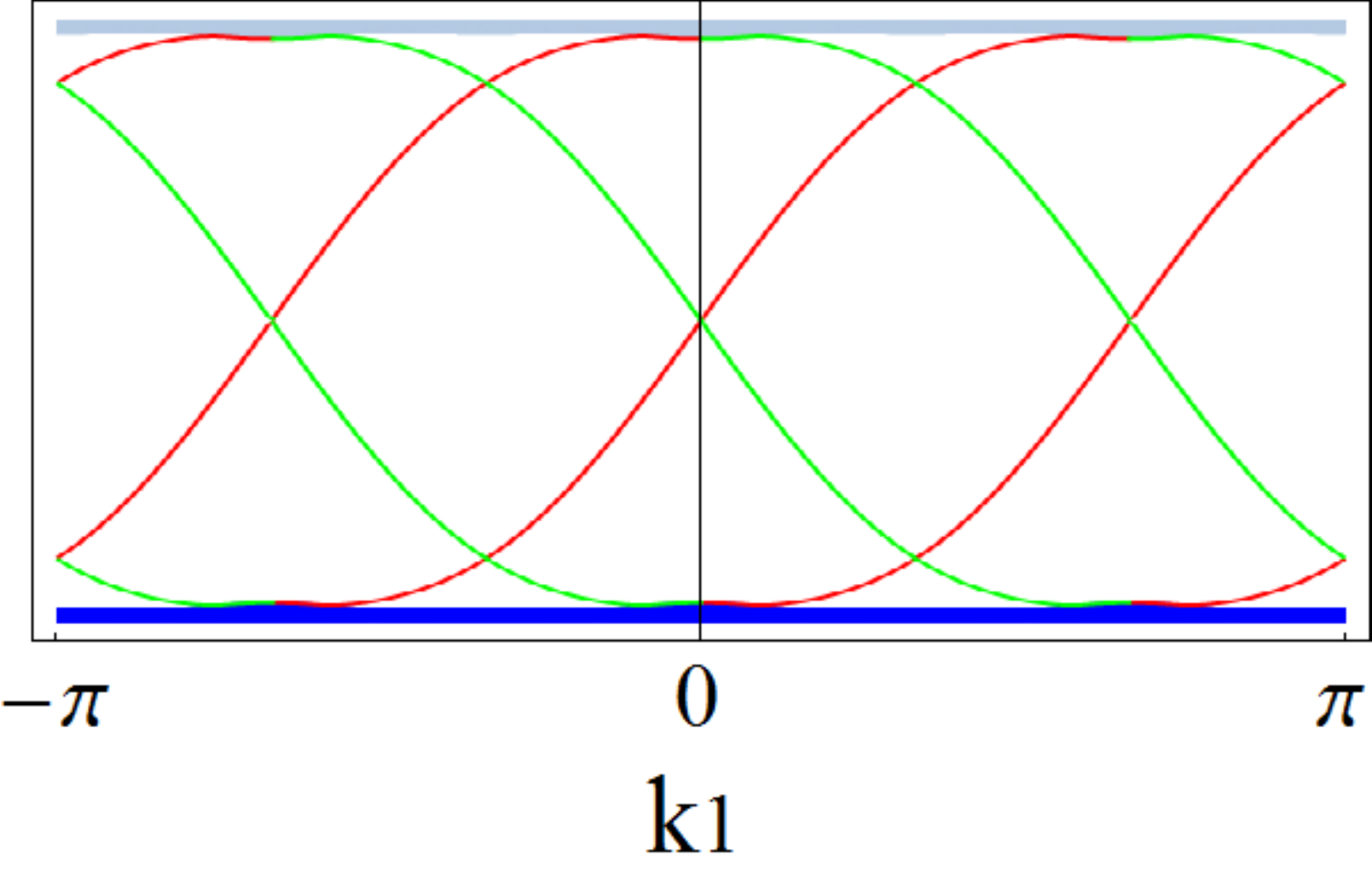}\label{fig:modelC3_edge}}
\caption{(Color online) The two band model with Chern number $C=3$ on a triangular lattice. (a) Lattice structure and NN (solid bond) and NNNN (dashed curve) hoppings. The end point of a link is red (blue) if it is connected to a $A$ ($B$) orbital. (b) Lattice structure and NNN hoppings. Arrows show the sign of the phases. 
The green disk indicates that each unit cell contains one lattice site and the green rectangle represents the enlarged unit cell. (c) Single-particle energy spectrum as a function of $k_{x}$ and $k_{y}$. (d) Chiral edge states of the two-orbital triangular lattice model, where $k_{1}=k_{x}/2+\sqrt{3} k_{y}/2$. Each band has Chern number $C=\pm 3$.}
\end{center}
\end{figure}

{\it Generalized Haldane model with Chern number $C=3$.} We first consider a triangular lattice model with two orbitals per site as shown in Fig.~\ref{fig:modelC3_NN}. Here, every site is colored in red and blue (dark gray and light gray) to present the two orbitals $A$ and $B$. The nearest-neighbor (NN) and next-next-nearest-neighbor (NNNN) hoppings are shown in Fig.~\ref{fig:modelC3_NN}, while Fig.~\ref{fig:modelC3_NNN} shows the next-nearest-neighbor (NNN) hoppings. The inter-orbital NN and NNNN (intra-orbital NNN) hoppings are illustrated by gradient-colored (even-colored) links. The end point of a link is red (blue) if it is connected to the $A$ ($B$) orbital. The Hamiltonian of this model is
\begin{align}
H^{[C=3]} =& -t_{1}\sum_{\langle i,j \rangle } A_{i}^{\dagger} B_{j} -t_{2} \sum_{\langle \langle i,j \rangle \rangle } e^{i \phi_{ij}} \left( A_{i}^{\dagger}A_{j}+B_{i}^{\dagger}B_{j} \right) \notag \\
& -t_{3} \sum_{\langle \langle \langle i,j \rangle \rangle \rangle } A_{i}^{\dagger} B_{j} +\mathrm{H.c.},
\label{HamiltonianC3}
\end{align}
where $A_{i}^{\dagger}$ and $B_{i}^{\dagger}$ are the fermion creation operators of the two orbitals at site $i$. The NN, NNN, and NNNN bonds are represented by $\langle i,j \rangle$, $\langle\langle i,j \rangle \rangle$, and $\langle \langle \langle i,j \rangle \rangle \rangle$. The phase factors in the NNN hopping terms are $\phi_{ij}=\pm \pi/2$ with the sign determined by the direction of the arrow~\cite{SI}.

In this model, each unit cell contains two orbitals, as shown by the green disk in Figs.~\ref{fig:modelC3_NN} and \ref{fig:modelC3_NNN}, and therefore the model has two bands.
In order to reach a flat band, we adopt $t_{1}=1$, $t_{2}=0.39$, and $t_{3}=-0.34$. The NNN hopping is purely imaginary, and hence it is easier to realize in optical lattices
or solid state materials through spin orbit couplings. The band structure of this model is shown in Fig.~\ref{fig:modelC3_band}.
Due to the particle-hole symmetry at $\phi_{ij}=\pm \pi/2$, the top and bottom bands have the same bandwidth, which is about $1/30$ of the band gap.
Each of the two bands carries Chern number $\pm 3$~\cite{chernnumber}. The nontrivial topological structure can also be observed 
by computing the edge states of the system on a cylinder [Fig.~\ref{fig:modelC3_edge}].

\begin{figure}[tbp]
\begin{center}
\subfigure[]{\includegraphics[width=0.43\columnwidth]{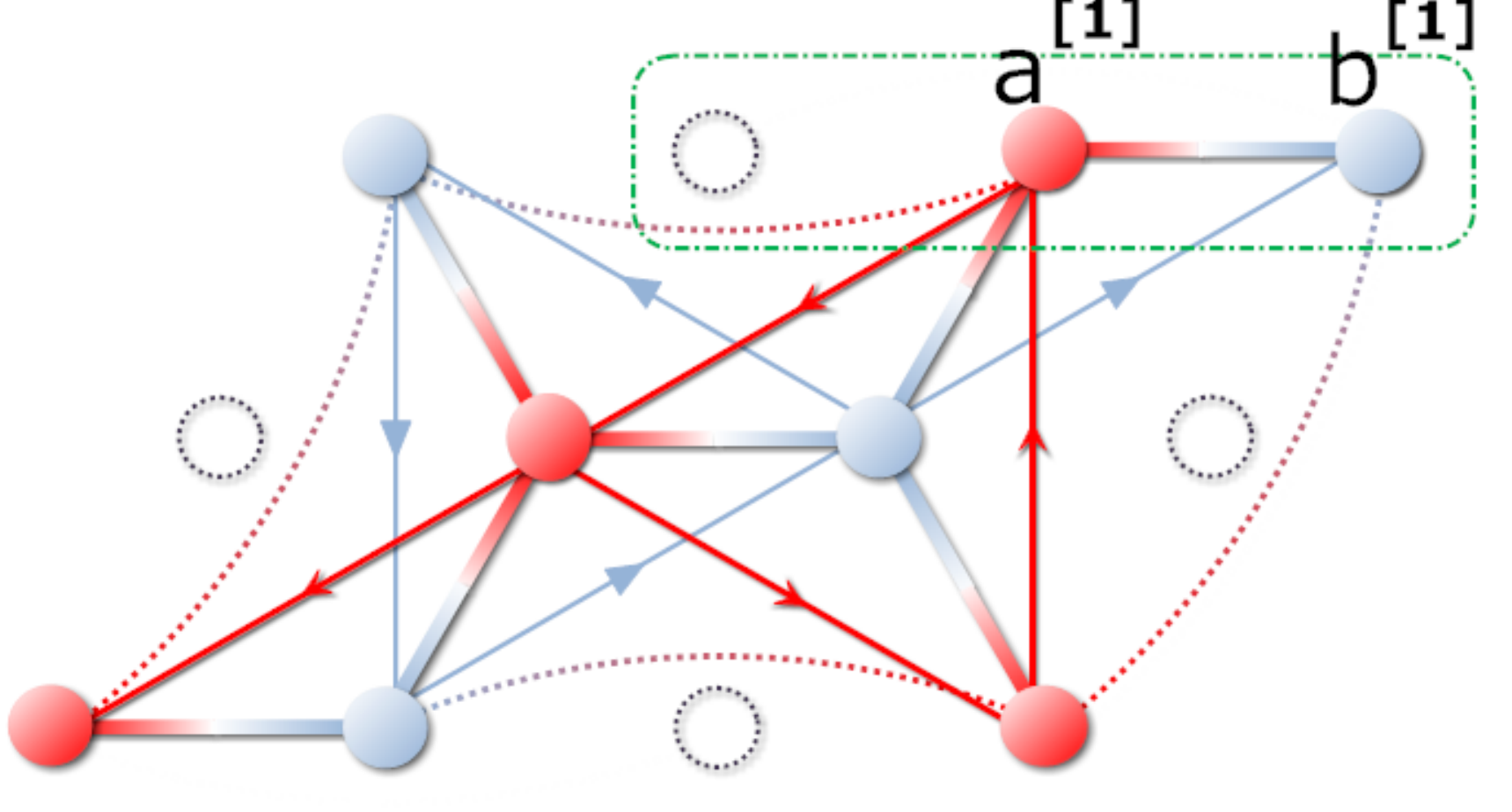}\label{fig:sepC3_layer1}}
\subfigure[]{\includegraphics[width=0.43\columnwidth]{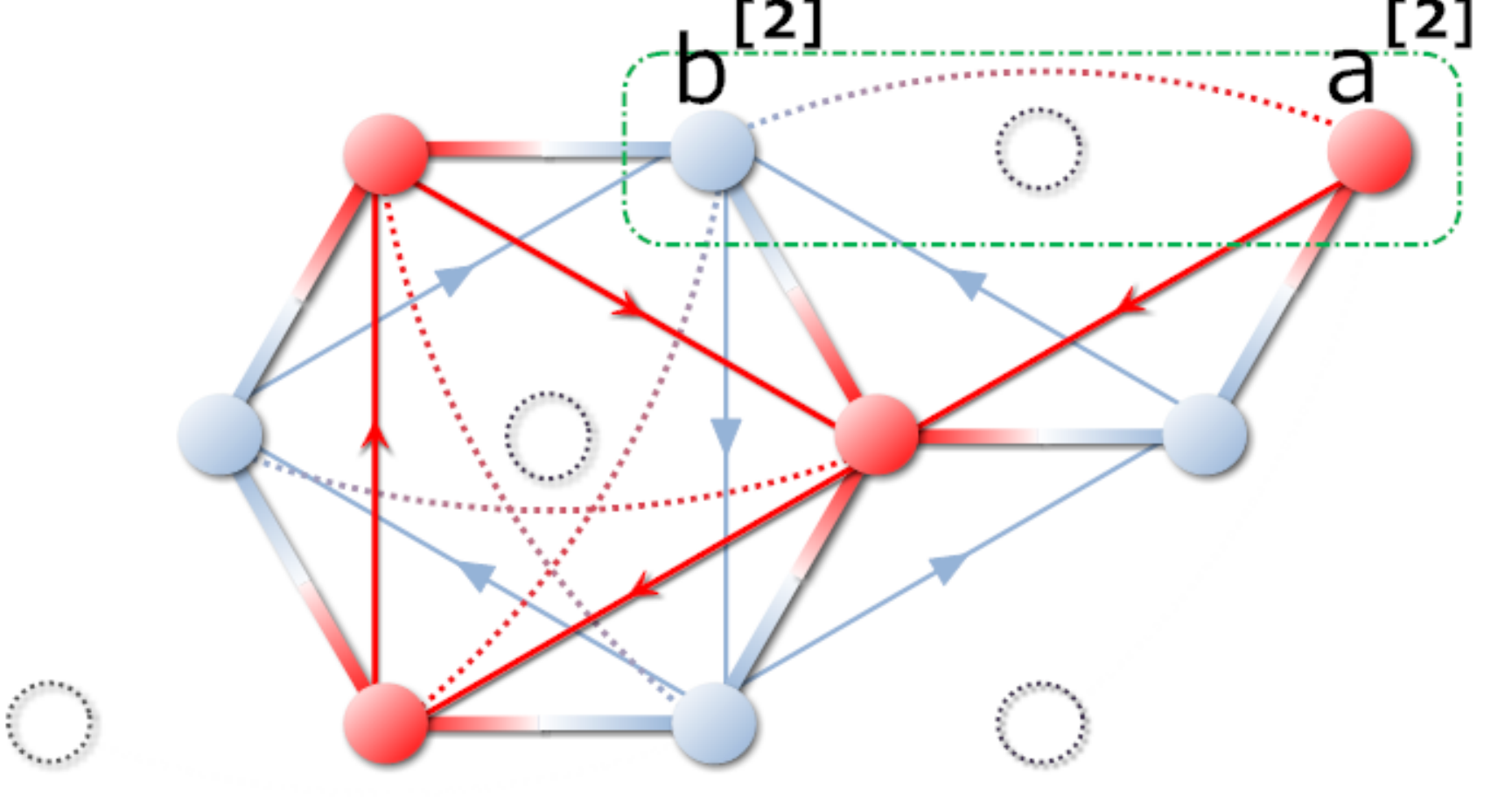}\label{fig:sepC3_layer2}}
\subfigure[]{\includegraphics[width=0.43\columnwidth]{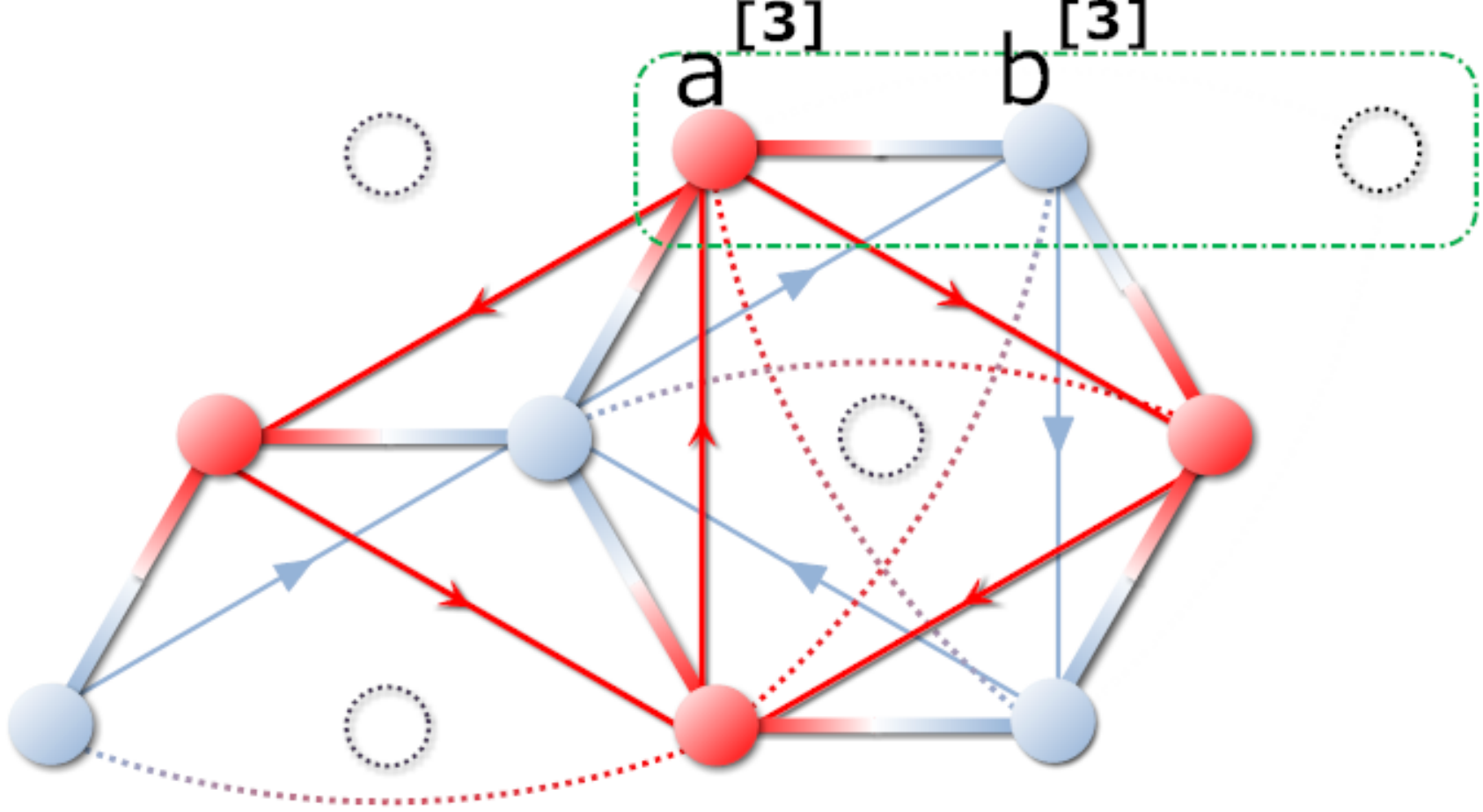}\label{fig:sepC3_layer3}}
\hspace*{0.05cm}
\subfigure[]{\includegraphics[width=0.4\columnwidth]{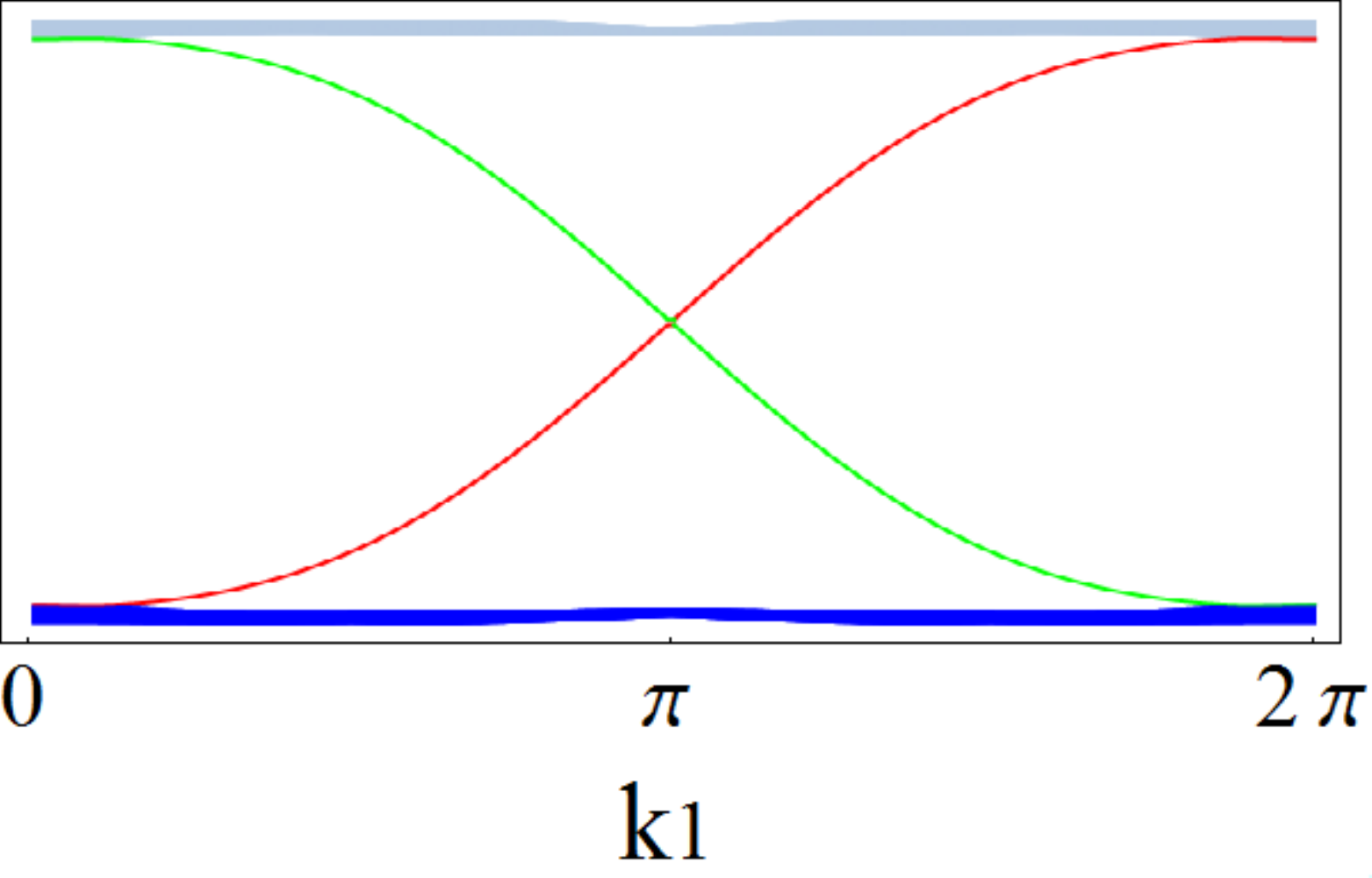}\label{fig:sepC3_edge}}
\end{center}
\caption{(Color online) (a)-(c) Tri-layer Haldane model in an ABC-type tri-layer structure. The green rectangle in each layer indicates the unit cell of the Haldane model. The dashed black circles mark the positions of some sites in the other two layers. (d) Chiral edge state of the Haldane model, where $k_{1}=k_{x}/2+\sqrt{3} k_{y}/2$.}
\end{figure}

\begin{figure*}[tbp]
\begin{center}
\subfigure[]{\includegraphics[width=0.19\textwidth]{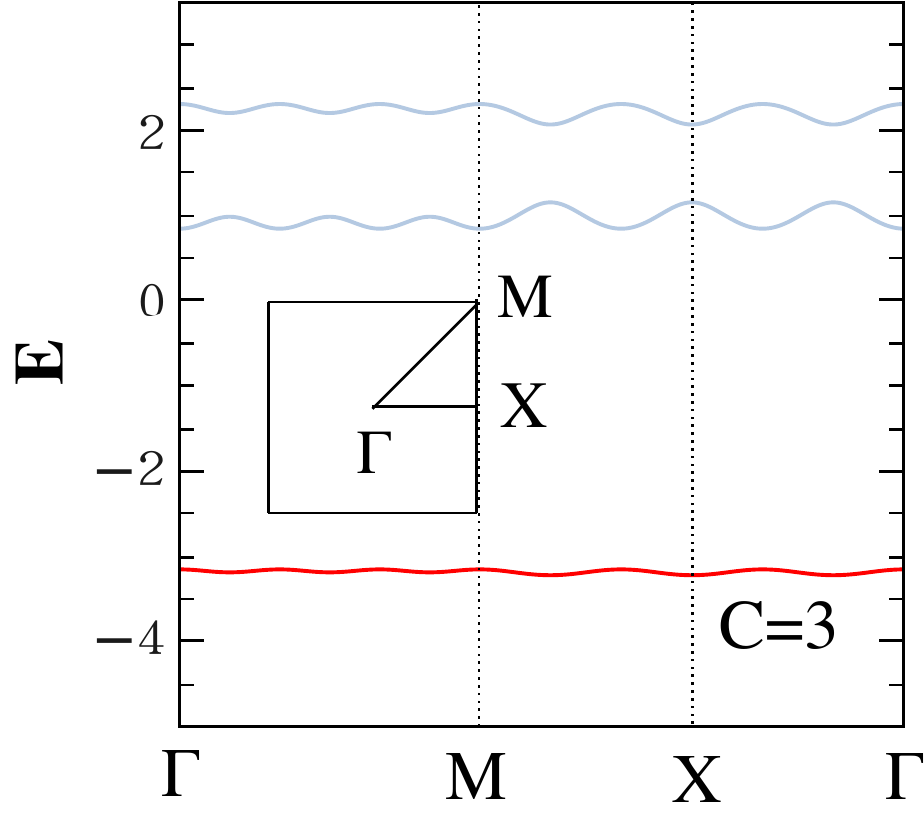}\label{fig:CN_P4C3}}
\subfigure[]{\includegraphics[width=0.19\textwidth]{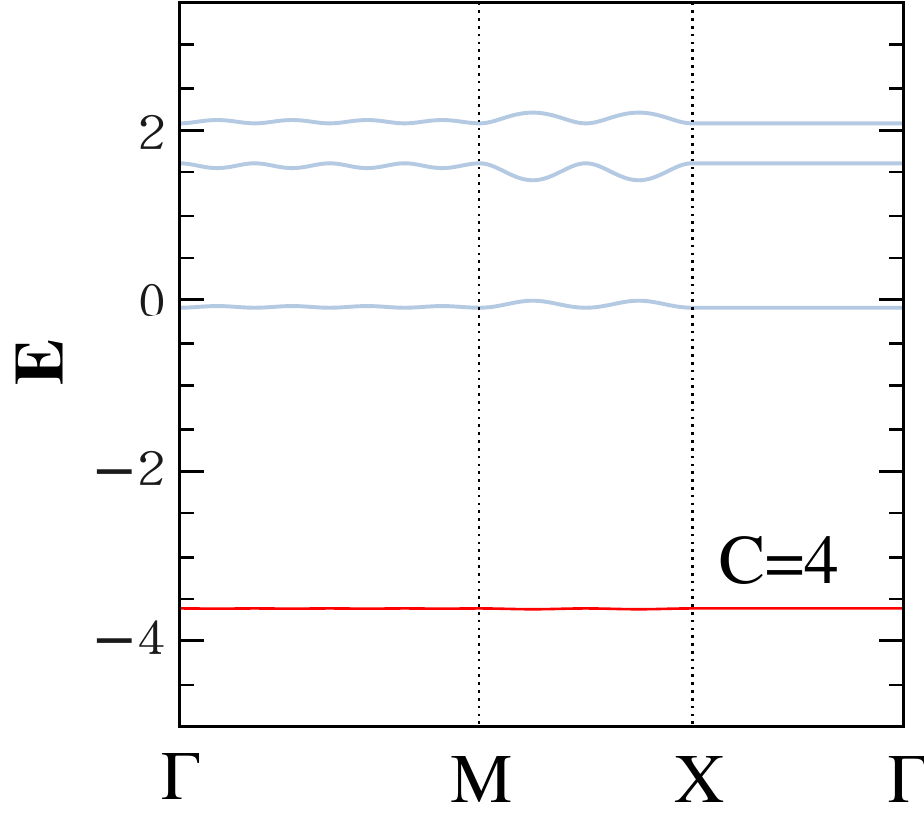}\label{fig:CN_P4C4}}
\subfigure[]{\includegraphics[width=0.19\textwidth]{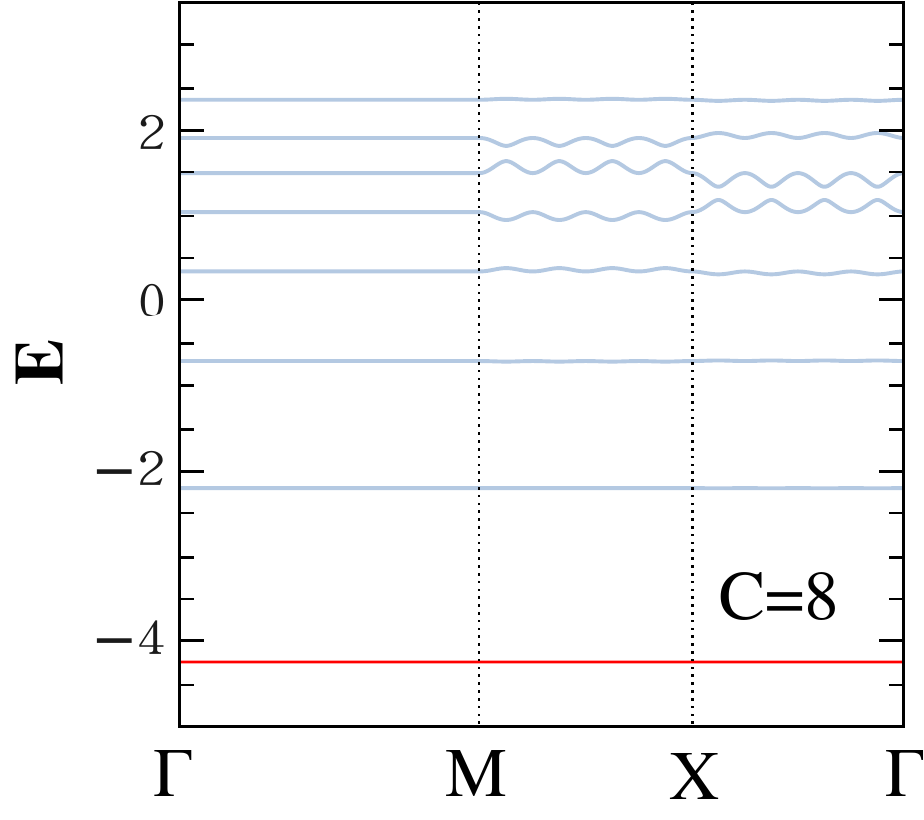}\label{fig:CN_P4C8}}
\subfigure[]{\includegraphics[width=0.19\textwidth]{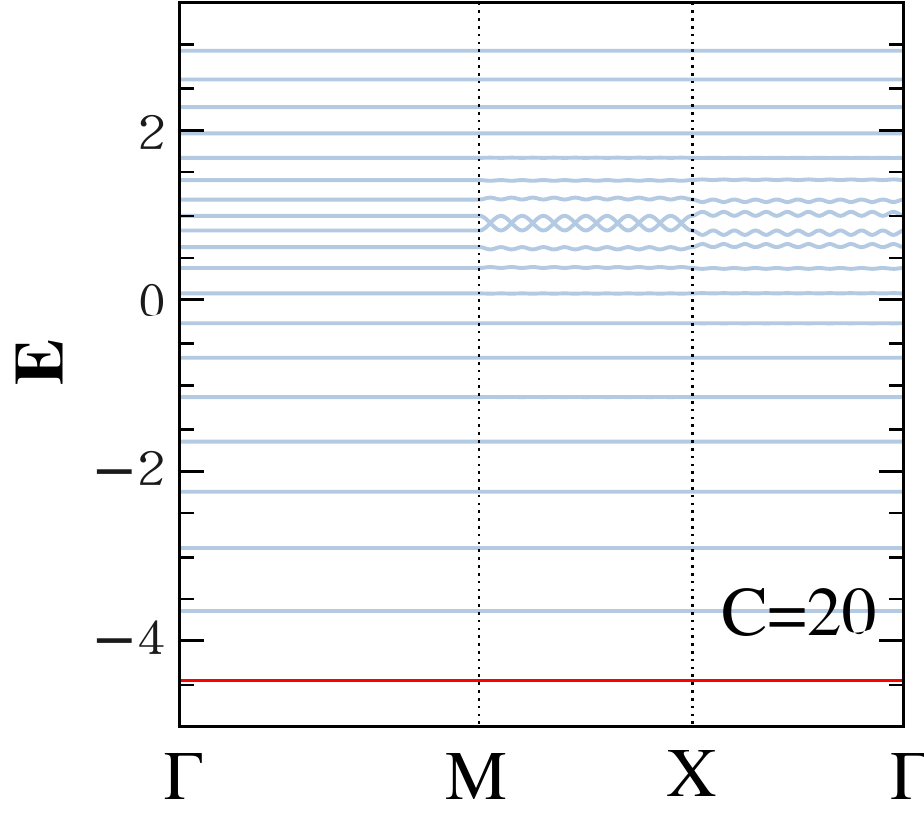}\label{fig:CN_P4C20}}
\subfigure[]{\includegraphics[width=0.19\textwidth]{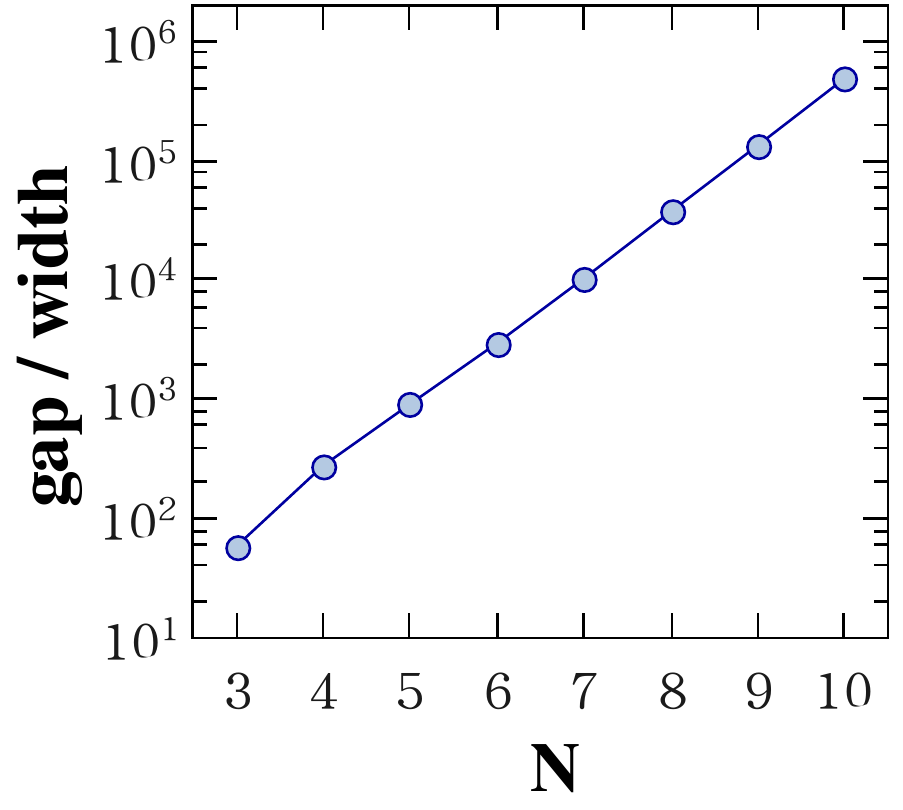}\label{fig:CN_flatness}}
\end{center}
\vspace{-0.2in}
\label{ChernN}
\caption{(Color online) (a)-(d) Dispersions of the $N$-orbital square lattice model along high symmetry directions for $N=3$, $4$, $8$, and $20$, where $t_{1}=1$, $t_{2}=-1/\sqrt{N}$, and $\phi=\pi/N$. In each case the bottom band (red line) is very flat with Chern number $C=N$. (e) The flatness ratio of the bottom flat band increases exponentially with the number of orbitals $N$ (the Chern number $N$).}
\vspace{-0.2in}
\end{figure*}

Here we provide an intuitive understanding for this $C=3$ model and demonstrate its connection to the model of Haldane~\cite{Haldane1988} and the ABC-type tri-layer graphene structure. We first consider Haldane's model on a honeycomb lattice. As shown in Fig.~\ref{fig:sepC3_layer1}, this lattice has two sites per unit cell,  $a^{[1]}$ and $b^{[1]}$, highlighted by the green rectangle. The primitive vectors for this lattice are $\vec{a}_{1,2}=(3a_{0}/2,\pm \sqrt{3}a_{0}/2)$ with $a_{0}$ being the length of the NN bond.
Now we stack three layers of this model in an ABC-type tri-layer pattern as shown in  Figs.~\ref{fig:sepC3_layer1}-\ref{fig:sepC3_layer3}.  In the absence of interlayer hopping, all three layers share the same Hamiltonian
\begin{align}
H_{\mathrm{Haldane}}^{[l]}&= -t_{1} \sum_{\langle i,j \rangle } a_{i}^{[l]\dagger} b_{j}^{[l]} -t_{3} \sum_{\langle \langle \langle i,j \rangle \rangle \rangle } a_{i}^{[l]\dagger} b_{j}^{[l]} \notag \\
& -t_{2} \sum_{\langle \langle i,j \rangle \rangle } e^{i \phi_{ij}} \left( a_{i}^{[l]\dagger}a_{j}^{[l]}+b_{i}^{[l]\dagger}b_{j}^{[l]} \right) +\mathrm{H.c.},
\end{align}
where $l=1,2,3$ denotes the different layers. It is well-known that the lowest band of the Haldane model carries a Chern number $C=1$ and there exists one chiral edge mode at each edge as shown in Fig. \ref{fig:sepC3_edge}. Therefore, one naively expects the tri-layer structure to contain three degenerate topological bands with $C=1$. However, in addition to increasing the number of bands, the tri-layer structure also enhances the translational symmetry of the system. In the absence of inter-layer hopping, the tri-layer structure is invariant under the translation by the vector $(a_{0},0)=(\vec{a}_1+\vec{a}_2)/3$, which is a fraction of the lattice vector of the original honeycomb lattice, if we also permute the layer indices $1\rightarrow 2$, $2\rightarrow 3$, and $3\rightarrow 1$. Due to such an enhancement of the translational symmetry, the area of a unit cell for the tri-layer system is \emph{reduced} by a factor of three containing only a single site with two orbitals. As a result, the Brillouin zone is \emph{enlarged} by three times and thus
the three-fold degenerate $C=1$ bands form a single band with $C=3$. It is easy to check that this tri-layer model is identical to the triangular lattice model presented above, if we introduce a unitary transformation
\begin{eqnarray}
A_{i1}^{\dagger}&=&a^{[3]\dagger}, A_{i2}^{\dagger}=a^{[1]\dagger}, A_{i3}^{\dagger}=a^{[2]\dagger}, \nonumber\\
B_{i1}^{\dagger}&=&b^{[2]\dagger}, B_{i2}^{\dagger}=b^{[3]\dagger}, B_{i3}^{\dagger}=b^{[1]\dagger},
\label{transformationC3}
\end{eqnarray}
where the subscripts $i1$, $i2$ and $i3$ mark different lattice sites.

\begin{figure}[tbp]
\begin{center}
(a)\includegraphics[width=0.33\columnwidth]{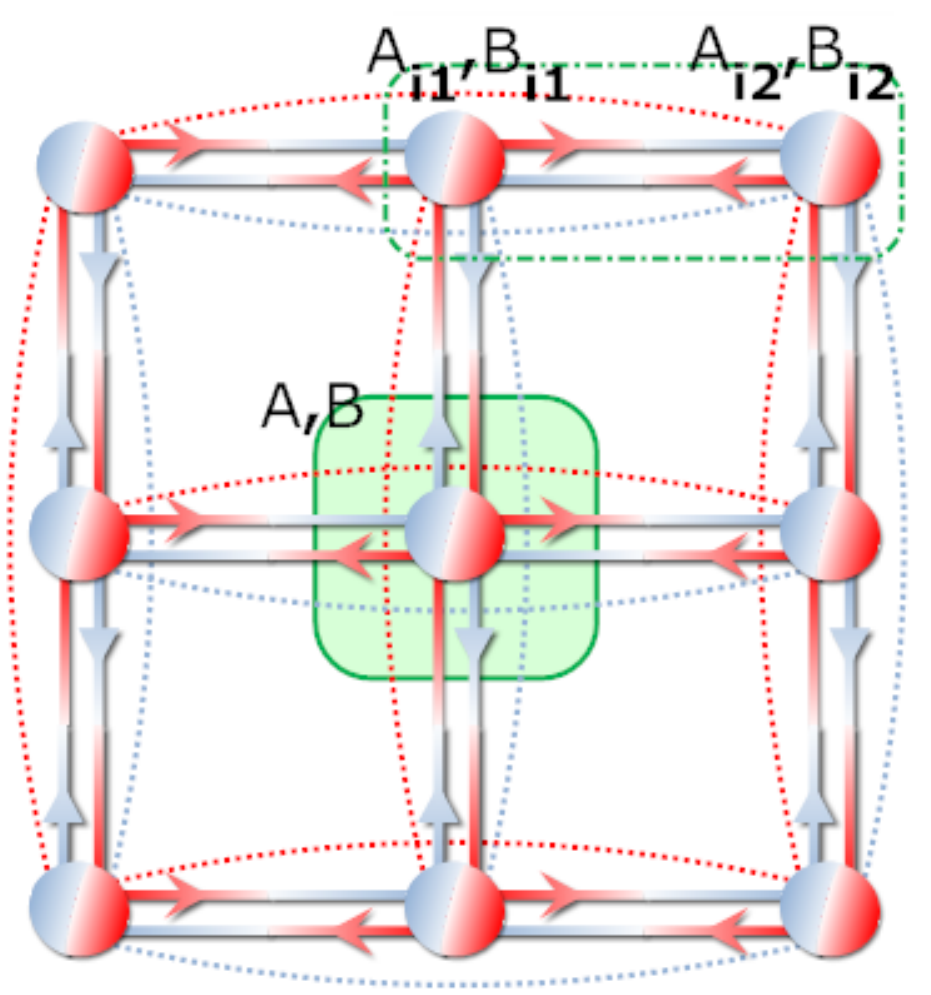}
\hspace*{0.3cm}
(b)\includegraphics[width=0.33\columnwidth]{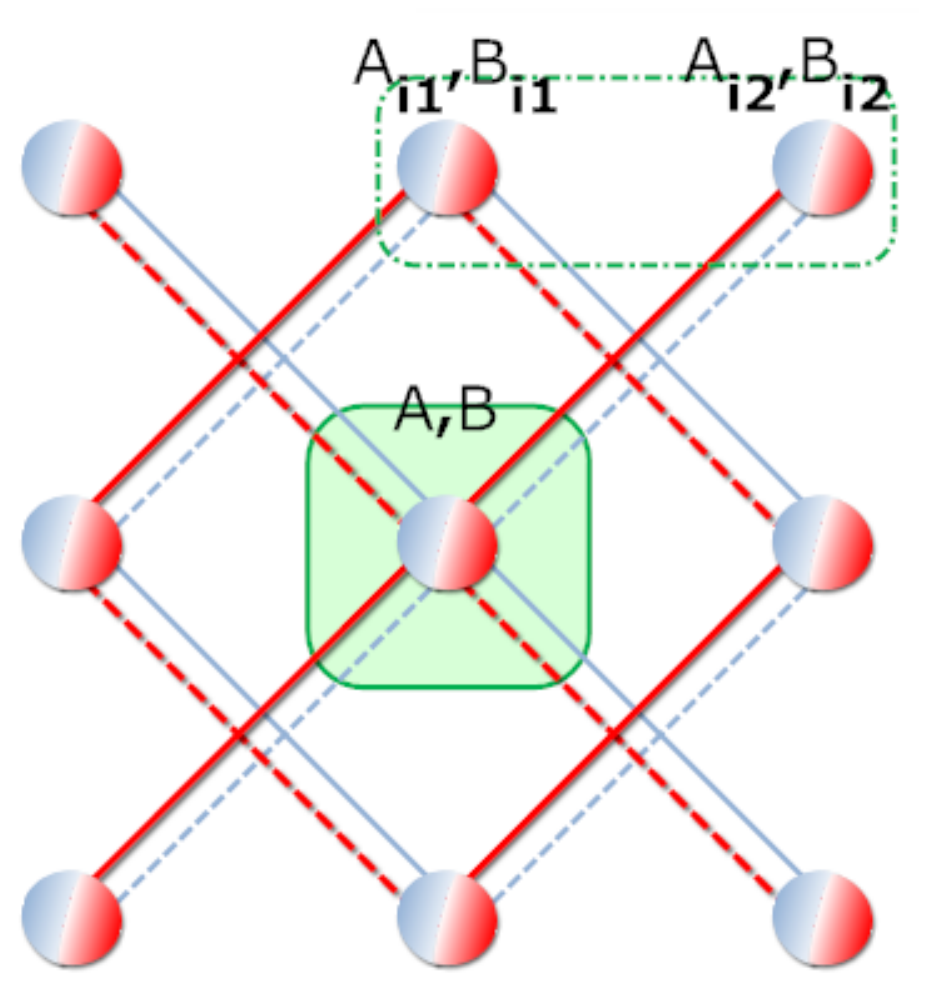}
\includegraphics[width=1.0\columnwidth]{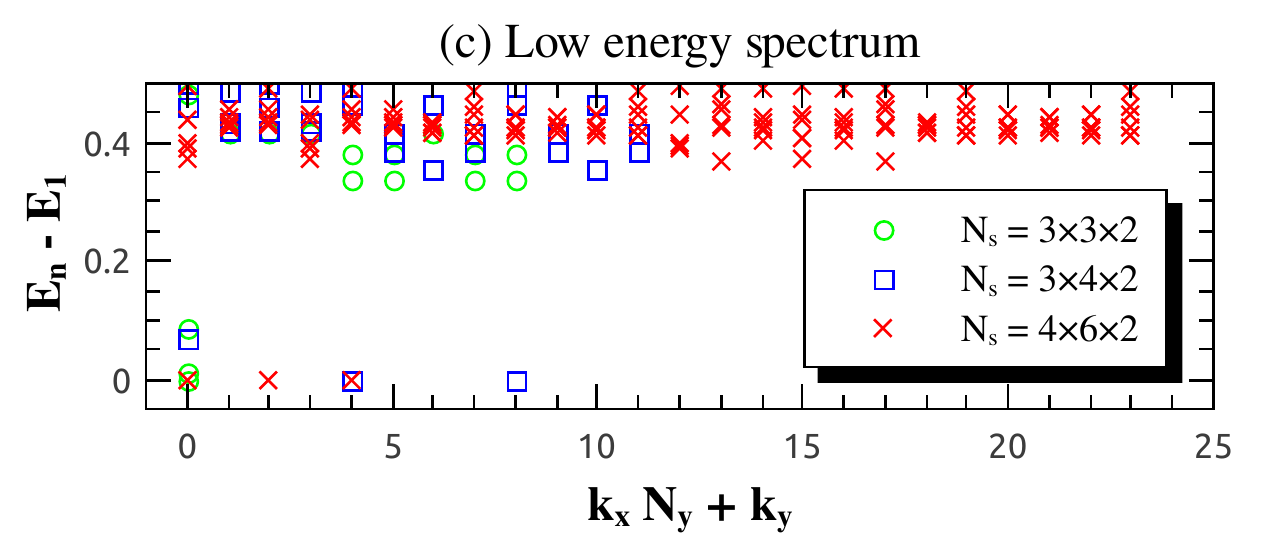}
\includegraphics[width=1.0\columnwidth]{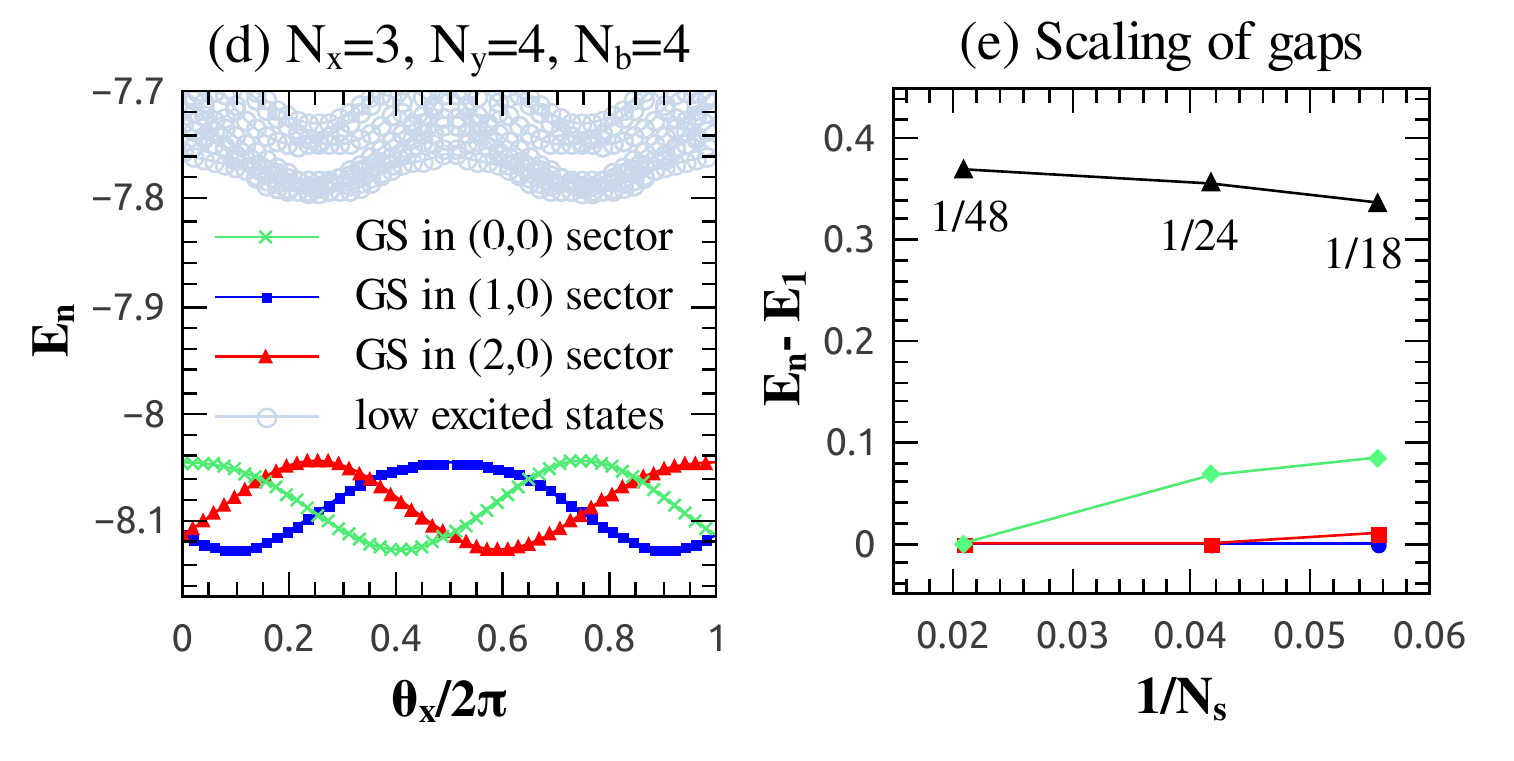}
\end{center}
\vspace{-0.1in}
\caption{(Color online) (a)-(b) The two-orbital Chern number two model on a square lattice. (a) NN and NNNN hoppings. (b) NNN hoppings. (c)-(e) The $1/3$ bosonic FQHE. (c) Low energy spectrum $E_{n}-E_{1}$ versus the momentum $k_{x} N_{y}+k_{y}$ for three lattice sizes $N_{s}=18$, $24$, and $48$. (d) Low energy spectrum as a function of boundary phase $\theta_{x}$ at a fixed $\theta_{y}=0$ at $1/3$ filling. (e) Spectrum gaps versus $1/N_{s}$ for three lattice sizes.}
\label{fig:boson221}
\vspace{-0.1in}
\end{figure}

{\it Generalization to Chern number $N$ topological flat bands.} The same idea can also be used to generate TFBs with arbitrary Chern numbers.
Without loss of generality, we consider a square lattice model. We first demonstrate this construction using the simplest case with Chern number $C=2$. The lattice and hoppings
are shown in Figs.~\ref{fig:boson221}(a) and ~\ref{fig:boson221}(b), where each site contains two orbitals ($A$ and $B$) and the arrows on NN bonds mark the directions of positive phase hopping.
The Hamiltonian of this model is
\begin{align}
H^{[C=2]} &= t_{1}\sum_{\langle i,j \rangle } e^{i \phi_{ij}} A_{i}^{\dagger} B_{j} +\sum_{\langle \langle i,j \rangle \rangle } t'_{ij} \left( A_{i}^{\dagger}A_{j}+B_{i}^{\dagger}B_{j} \right) \notag \\
& +t_{3} \sum_{\langle \langle \langle i,j \rangle \rangle \rangle } \left( A_{i}^{\dagger}A_{j}+B_{i}^{\dagger}B_{j} \right) +\mathrm{H.c.},
\label{HChern2}
\end{align}
The NNN hopping amplitudes $t'_{ij}$ are $t_{2}$ ($-t_{2}$) along the solid (dashed) lines in Fig.~\ref{fig:boson221}(b). Here we adopt the parameters $t_{1}=1$, $t_{2}=1/(2+\sqrt{2})$, $t_{3}=1/(2+2\sqrt{2})$, and $\phi_{ij}=\pm \pi/4$~\cite{SI}. In this two-band model, the bottom band carries Chern number $C=2$ and the flatness ratio is about 30. 
This Chern number two model can be constructed from a bilayer version of the checkerboard lattice model first proposed in Ref.~\cite{sun} (See SI for details).
Similar to the tri-layer case presented above, this bi-layer structure also enhances the translational symmetry and reduces the size of the unit cell by a factor of two.

The generalization to flat band models carrying an arbitrary Chern number $C=N$ is straightforward. Consider a square-lattice model with $N$ orbitals per site. For $N \geq 3$, we only need NN and NNN hoppings. The real space Hamiltonian is
\begin{align}
H^{[C=N]} & = \sum_{i,j}\sum_{l=1}^{N}\left\{ t_{1}\left(C_{i+1,j}^{[l+1]\dagger}+e^{i2l\phi}C_{i,j-1}^{[l+1]\dagger}\right)C_{i,j}^{[l]} \right. \notag \\
 &  +t_{2}\left[e^{-i\left(2l-1\right)\phi}C_{i+1,j+1}^{[l]\dagger}+e^{i\left(2l-1\right)\phi}C_{i-1,j-1}^{[l]\dagger}\right. \notag \\
&  \left.\left.+e^{i\left(2l+1\right)\phi}C_{i+1,j-1}^{[l+2]\dagger}\right]C_{i,j}^{[l]}+\mathrm{H.c.}\right\},
\end{align}
The $k$-space Hamiltonian is shown in the Supplemental Materials.

For $N=3$, the flatness ratio is maximized ($\simeq 58$) at $t_{2}/t_{1}=-1/\sqrt{3}$ and $\phi=\pi/3$ [Fig.~\ref{fig:CN_P4C3}]. For the generic case with $C=N$, we adopt $t_{1}=1$, $t_{2}=-1/\sqrt{N}$, and $\phi=\pi/N$ for the $N$-orbital model. With these values, the flat band is always located at the bottom and the Chern number is $C=N$. In the absence of any band crossings, one of the higher bands carries Chern number $C=-(N-1)N$, while each of the other $(N-1)$ bands carries Chern number $C=N$. The band structures of the $C=3$, $4$, $8$, and $20$ cases are shown in Figs.~\ref{fig:CN_P4C3}-~\ref{fig:CN_P4C20}. We see that the bottom flatband becomes more and more flat with an increasing $N$. 
As shown in Fig.~\ref{fig:CN_flatness}, the flatness ratio grows exponentially with the Chern number $N$. The condition $t_{2}=-1/\sqrt{N}$ can be further relaxed for larger $N$ systems. In particular, it suffices to only include NN hoppings when $N \geq 5$.

The $N$-orbital $C=N$ model can be constructed using a $N$-layer structure. All the $N$ layers are exactly the same up to lattice translations. Each layer is equivalent to a square lattice subjected to a uniform magnetic field. The complex phases of the hopping matrix elements are arranged in such a way that each square encircles a $\phi=2\pi/N$ flux. Due to the flux structure, the (magnetic) unit cell for each layer contains $N$ sites and the lowest band of each layer carries Chern number $C=1$. If we naively put the $N$ layers together, the unit cell has $N^2$ sites with the lowest bands $N$-fold degenerate. 
However, similar to the models presented above, this $N$-layer structure has a higher translational symmetry and reduces the size of the (magnetic) unit cell by $N$ times, if we introduce unitary transformations (simply a permutation of $N$ orbits) similar to Eqs.~\eqref{transformationC3}. As a result, the Brillouin zone is enlarged $N$ times and thus the $N$-fold degenerate $C=1$ bands form a single band with $C=N$. Here we emphasize that, although each layer is exposed to a magnetic field, the combined Chern number $N$ model is realized without an external magnetic field (with zero magnetic flux per unit cell and the translational symmetry is recovered).

{\it $1/3$ bosonic FQH state for the $C=2$ model.} Similar to $C=1$ TFB models, various FQH states can be stabilized in these high Chern number systems once proper interactions are introduced. As an example, we fill the the $C=2$ flatband model [Eq.~\eqref{HChern2}] with hard-core bosons at the filling fraction $1/3$. For a system with $N_{x} \times N_{y}$ unit cells, we have $N_{x}N_{y}=3N_{b}$, where $N_{b}$ is the number of hard-core bosons. The hard-core condition (no more than one particle is allowed per site) corresponds to infinite onsite repulsions: $\lim_{U \rightarrow \infty} U \sum_{i} (A_{i}^{\dagger} A_{i}+ B_{i}^{\dagger} B_{i})(A_{i}^{\dagger} A_{i}+ B_{i}^{\dagger} B_{i}-1)$.

In Fig.~\ref{fig:boson221} (c), we show the low-energy spectra as a function of momentum $k_{x}N_{y}+k_{y}$ for different lattices under periodic boundary conditions. Here, a sizable spectrum gap can be observed, separating the ground state manifold (GSM) from the other excited states and indicating the emergence of an incompressible state. As a definite evidence for the emergence of a FQH state, we also compute the Chern number, which is $2/3$ for these three-fold nearly-degenerate ground states. 
The scaling of the spectrum gaps is shown in Fig.~\ref{fig:boson221}(e) and in Fig.~\ref{fig:boson221} (d), we show the low energy spectra as a function of twist boundary phase along the $x$ direction with $N_{x}=3$ and $N_{y}=4$. As the phase parameter $\theta_{x}$ changes by $2 \pi$, $4 \pi$, etc., the system goes from one ground state to another, and it comes back to the original ground state after $\theta_{x}$ varies by $6 \pi$. According to the double layer picture for our $C=2$ model, the most possible candidate for such a FQH state is the double layer Halperin (221) state. The $C=2$ model is the simplest model and has the minimal flatness ratio in our systematic construction. Therefore, we believe FQH states will exist in any of our $C>2$ model.



{\it Conclusion.} Before concluding we mention that bilayer FQHE is well-known in the 2D LL systems, both experimentally and theoretically~\cite{Peterson2010a,Peterson2010b,He1993}. 
Multilayer QHE or FQHE (beyond bilayers) has not been considered much in the continuum systems in contrast to our current work where we find that $N$-layer QHE with a Chern number $C=N$ should be generic in lattice TFB systems. 
This is potentially an important finding, establishing the TFB lattice FQHE to be a much more general theoretical concept than the ordinary continuum LL-based FQHE so far realized in nature in high magnetic fields.

In conclusion, we propose a generic scheme to construct TFB models with arbitrary Chern number. These new models could be a starting point for many unknown Abelian or non-Abelian FQH states with higher Chern numbers ($>1$) which are generically unknown in the continuum 2D FQHE. A more detailed study of these models with interactions is beyond the scope of this Rapid Communication and will be discussed in our future works.

{\it Note added}.
After the completion of our work, we notice that in another recent work~\cite{Trescher2012}, a model with high Chern number bands was discovered. However, both the band gap and the flatness ratio in this model vanish as the Chern number increases. After the submission of the present Letter, another work \cite{Sterdyniak2012} observed FQH states in our $C=3$ triangular lattice model, and $C=4,5$ square lattice models.

{\it Acknowledgments}.
This work is supported by DARPA-QuEST, AFOSR MURI, and JQI-NSF-PFC. ZCG is partially supported by US NSF Grant No. NSFPHY05-51164. We thank D. N. Sheng for very helpful discussions on the numerical technique.

\begin{appendix}
\begin{widetext}
\section{Momentum space Hamiltonian and Berry curvature}

\begin{figure}[b]
\begin{center}
\subfigure[]{\includegraphics[width=0.32\columnwidth]{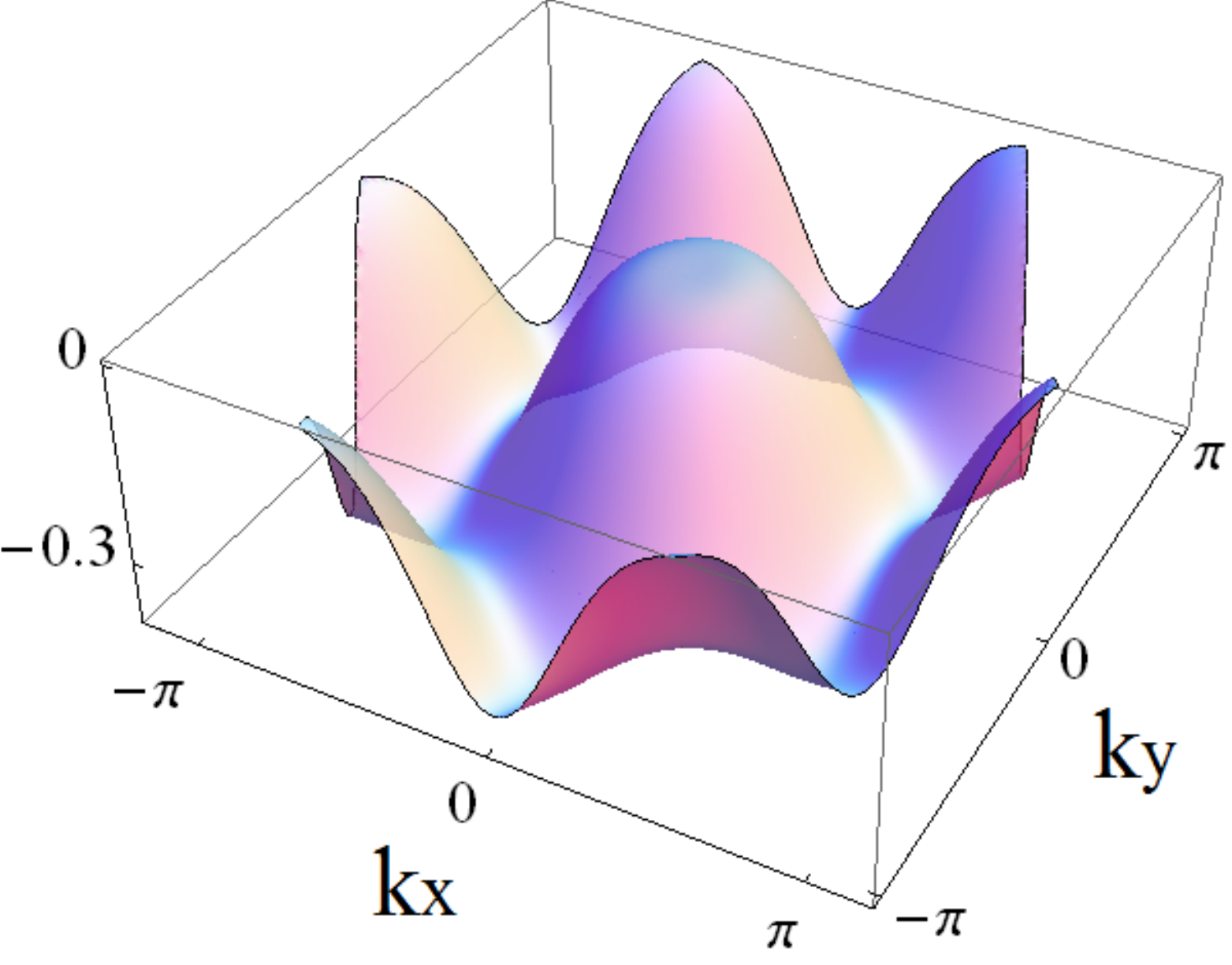}\label{fig:Berry_C3}}
\hspace{2cm}
\subfigure[]{\includegraphics[width=0.32\columnwidth]{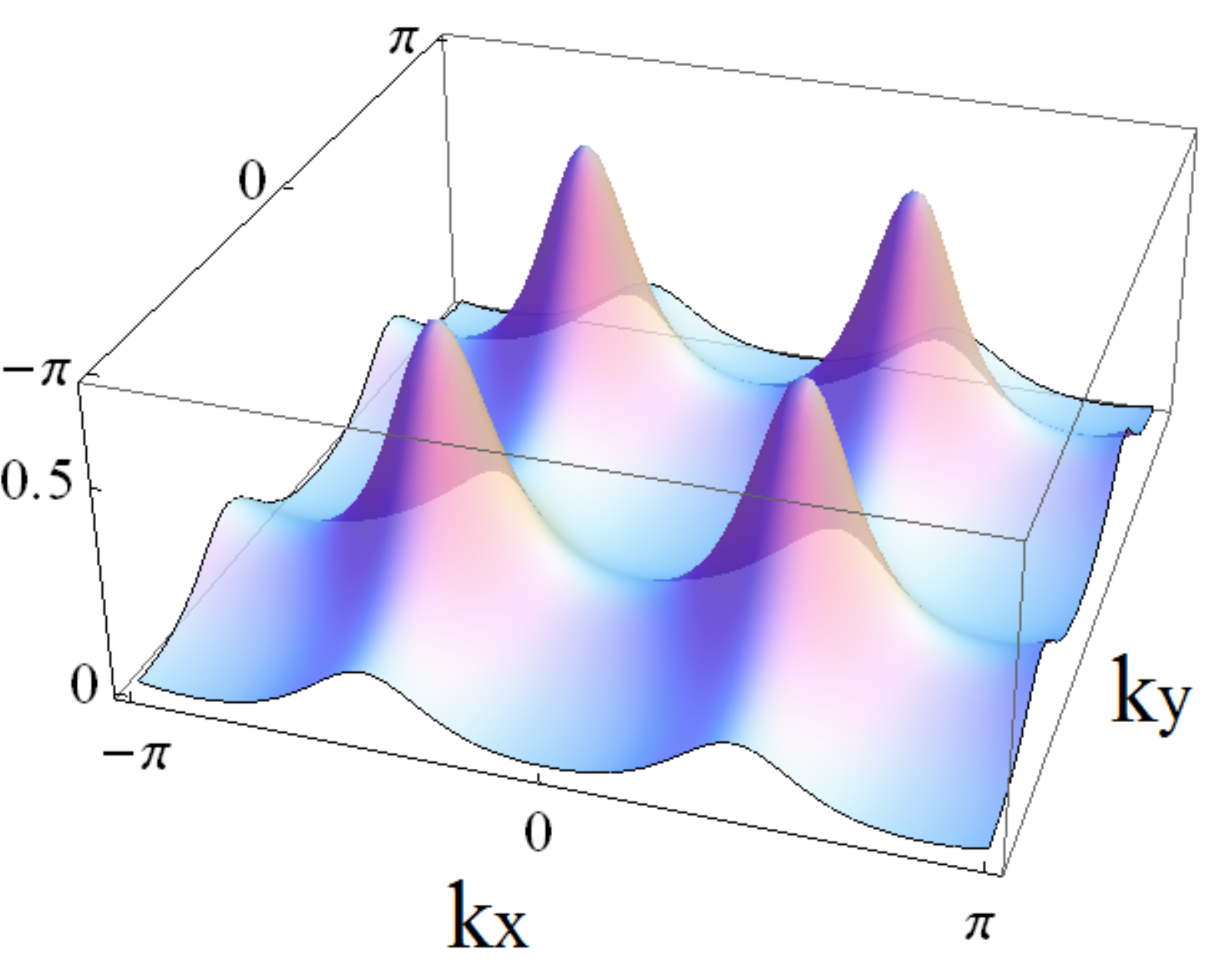}\label{fig:Berry_C2}}
\end{center}
\caption{(Color online) Distributions of the Berry curvature in momentum space for the flat bands in (a) the triangular lattice $C=3$ model and (b) the square lattice $C=2$ model.}
\end{figure}

Here we provide some final details for the Chern number $C=3$ and $C=2$ models proposed in this Letter. We will first give out their Hamiltonians in momentum space and then discuss the Berry curvature distribution in momentum space.
As shown in Figs.~1(a) and~1(b) in the main text,
the triangular lattice model is translational invariant along the lattice vectors $\mathbf{r}_{1}=(a_{0}/2, \sqrt{3} a_{0}/2)$ and $\mathbf{r}_{2}=(-a_{0}/2, \sqrt{3} a_{0}/2)$, where $a_{0}$ is the length of the NN bond. In momentum space, the tight-binding Hamiltonian is
\begin{align}
H^{[C=3]}=\sum_{\mathbf{k}} \Psi_{\mathbf{k}}^{\dagger} \mathcal{H}^{[C=3]} \Psi_{\mathbf{k}},
\end{align}
where $\Psi_{\mathbf{k}}=(A_{\mathbf{k}},B_{\mathbf{k}})$ is a two component spinor and $\mathcal{H}^{[C=3]}$ is a $2 \times 2$ matrix. Because the intra-orbital (NNN) hoppings are purely imaginary, $\mathcal{H}^{[C=3]}$ can be expressed in terms of three Pauli matrices,
\begin{align}
\mathcal{H}^{[C=3]}= \sum_{i} h_{i} \sigma_{i}
\end{align}
with $i=x$, $y$ and $z$. Here
\begin{align}
h_{x} =& -t_{1} \left[ \cos k_{1} +\cos k_{2} +\cos (k_{1}-k_{2}) \right]
 -t_{3} \left[ \cos 2 k_{1} + \cos 2 k_{2} +\cos (2 k_{1}-2 k_{2}) \right], \notag \\
h_{y} =& +t_{1} \left[ \sin k_{1} -\sin k_{2} -\sin (k_{1}-k_{2})  \right]
 -t_{3} \left[ \sin 2 k_{1} - \sin 2 k_{2} -\sin (2 k_{1}-2 k_{2}) \right], \notag \\
h_{z} =& -2 t_{2} \left[ \sin (k_{1}- 2 k_{2}) -\sin (2 k_{1} -k_{2}) +\sin (k_{1}+k_{2}) \right],
\end{align}
where $\mathbf{k}=(k_{x},k_{y})$, $k_{1}=\mathbf{k} \cdot \mathbf{r}_{1}$, and $k_{2}=\mathbf{k} \cdot \mathbf{r}_{2}$.


In momentum space, the Hamiltonian of the $C=2$ square lattice model
is given by
\begin{align}
H^{[C=2]}=\sum_{\mathbf{k}} \Psi_{\mathbf{k}}^{\dagger} \mathcal{H}^{[C=2]} \Psi_{\mathbf{k}},
\end{align}
where $\Psi_{\mathbf{k}}=(A_{\mathbf{k}},B_{\mathbf{k}})$ and
\begin{align}
\mathcal{H}^{[C=2]} &=2 t_{3} \left( \cos 2 k_{x}+\cos 2 k_{y} \right) \mathbb{I} +\sqrt{2} t_{1} \left( \cos k_{x} +\cos k_{y} \right) \sigma_{x} -\sqrt{2} t_{1} \left( \cos k_{x} -\cos k_{y}\right)\sigma_{y} -4t_{2} \sin k_{x} \sin k_{y} \sigma_{z}.
\end{align}

The distributions of the Berry curvature in momentum space for the triangular lattice $C=3$ model and the square lattice $C=2$ model are shown in Figs.~S\ref{fig:Berry_C3} and~S\ref{fig:Berry_C2}, respectively. In both models, the Berry curvature in momentum space shows no sharp feature.

\section{Flatness ratio}
In Fig.~\ref{fig:flatpara}, we show the contour plot of the ratio of band gap/bandwidth for the nearly flat band at different
$t_{2}$ and $t_{3}$ ($20\%$ away from the values we used in the Letter) with $t_{1}=1$ and $\phi_{ij}=\pm \pi/2$. As can be seen from the figure, in a wide region of the parameter space, this ratio remains large indicating a very flat band.

\begin{figure}[h]
\begin{center}
\includegraphics[width=0.35\textwidth]{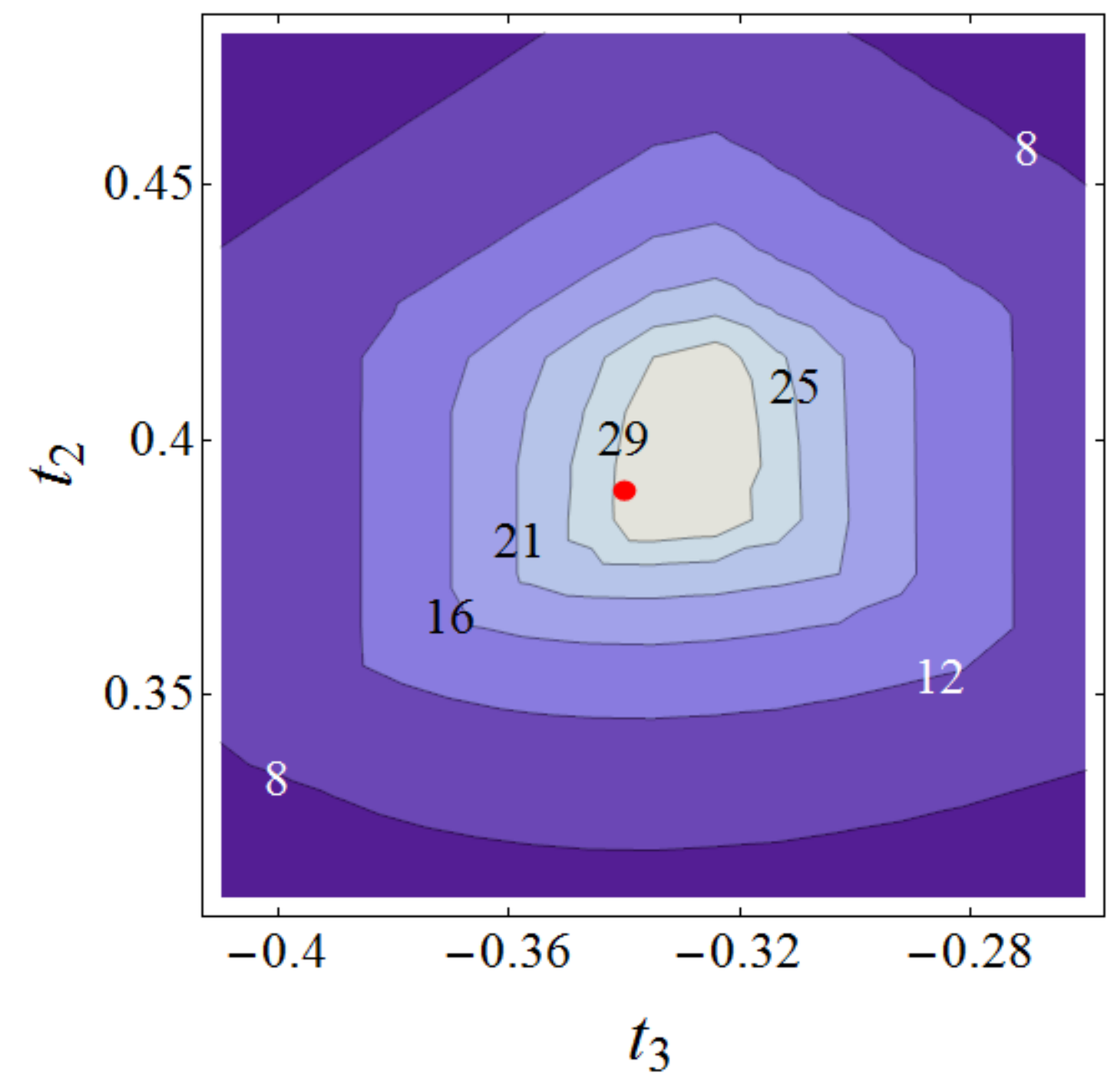}\label{fig:flat_parameter}
\end{center}
\caption{(Color online). The ratio of band gap/bandwidth for the flat band in the triangular lattice $C=3$ model. The two axes are the $NNN$ and $NNNN$ hopping strengths. The red dot at the center marks the parameters used in the Letter.}
\label{fig:flatpara}
\end{figure}

As for the square lattice $C=2$ model, its flatness ratio is identical to the checkerboard lattice model proposed in Ref.~\cite{sun}. Therefore, the corresponding contour plot is the same as Fig. 1.(b) in the Supplementary Materials of Ref.~\cite{sun}.

\section{$C=2$ square lattice model and bi-layer checkerboard lattice model}

\begin{figure}[tbp]
\begin{center}
\vskip 0.0cm \hspace*{-0.0cm}
\subfigure[]{\includegraphics[width=0.19\columnwidth]{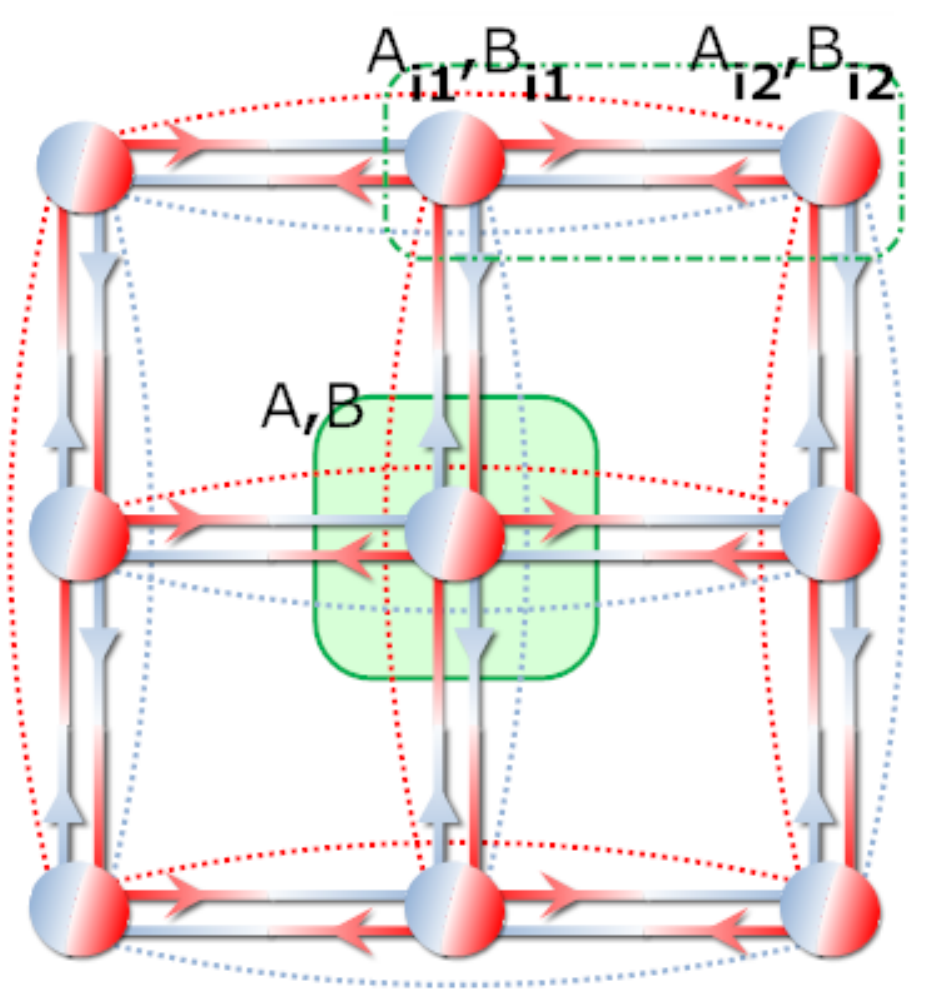}\label{fig:modelC2_NN}}
\hspace*{0.3cm}
\subfigure[]{\includegraphics[width=0.19\columnwidth]{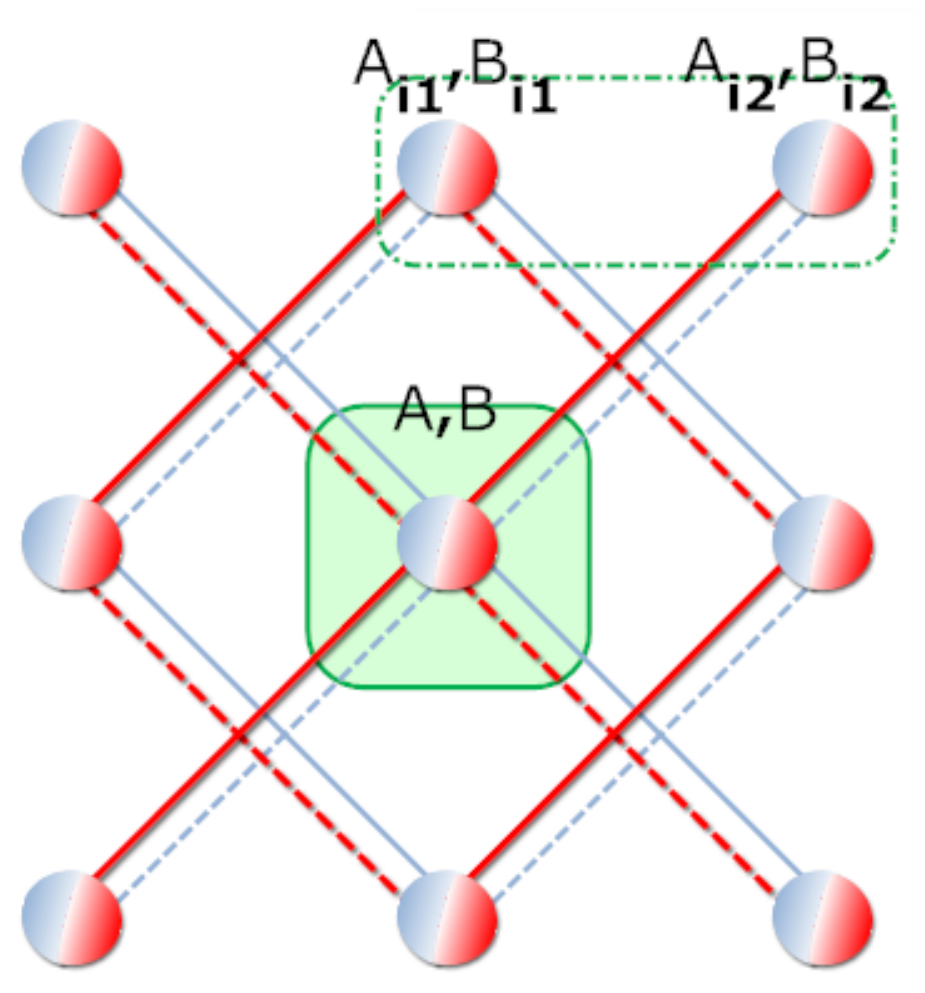}\label{fig:modelC2_NNN}}
\hspace*{0.3cm}
\subfigure[]{\includegraphics[width=0.22\columnwidth]{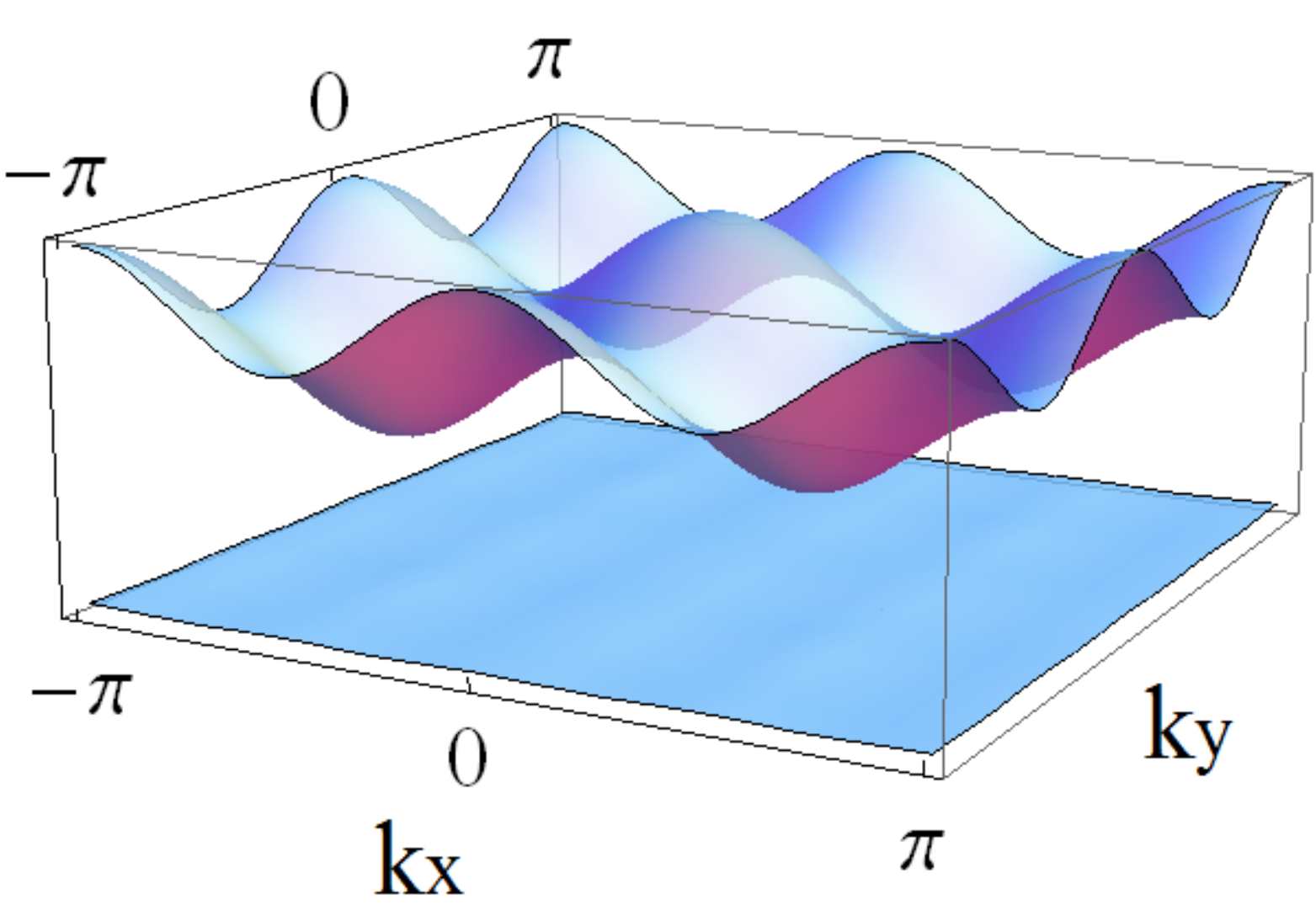}\label{fig:modelC2_band}}
\hspace*{0.3cm}
\subfigure[]{\includegraphics[width=0.22\columnwidth]{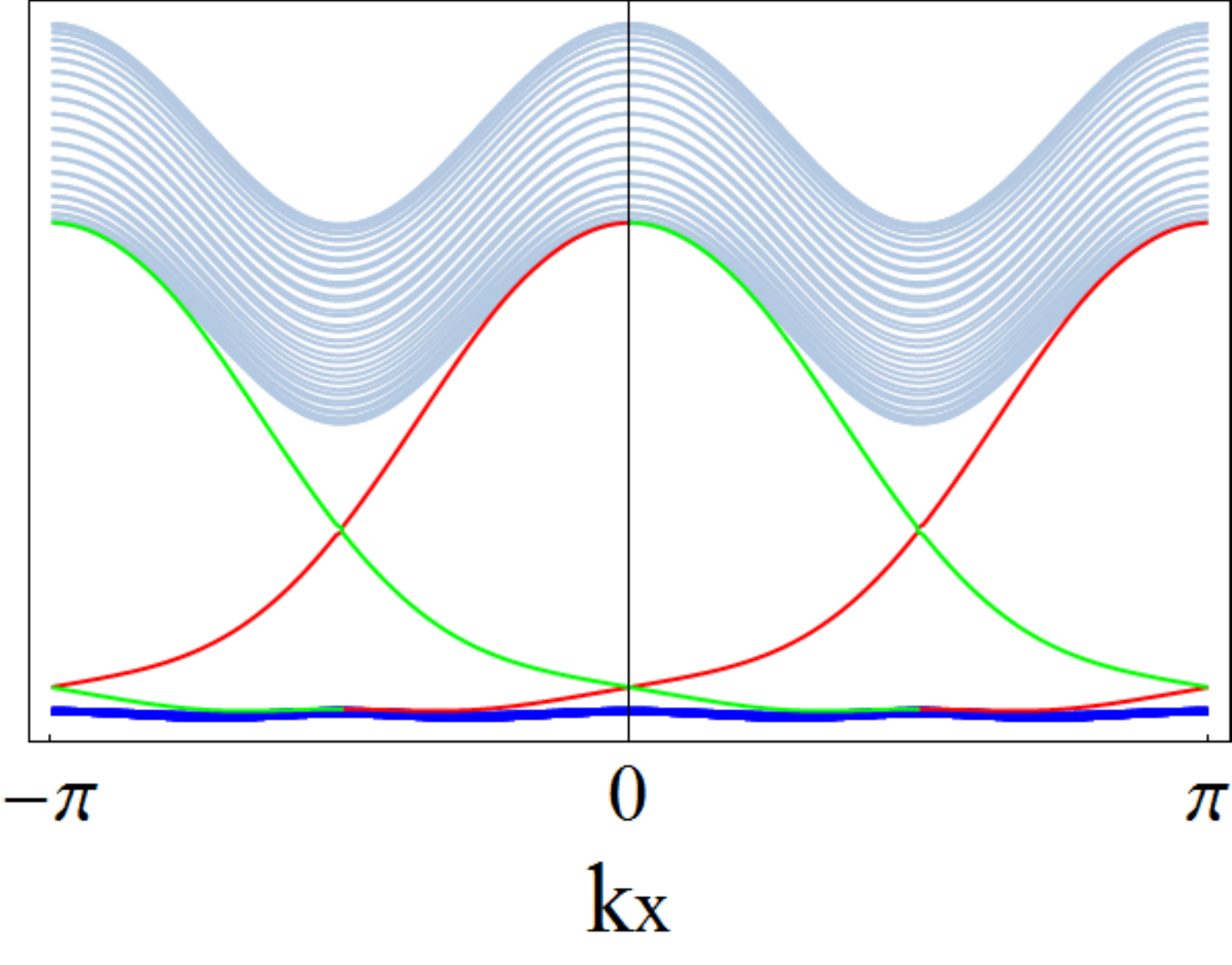}\label{fig:modelC2_edge}}
\subfigure[]{\includegraphics[width=0.20\columnwidth]{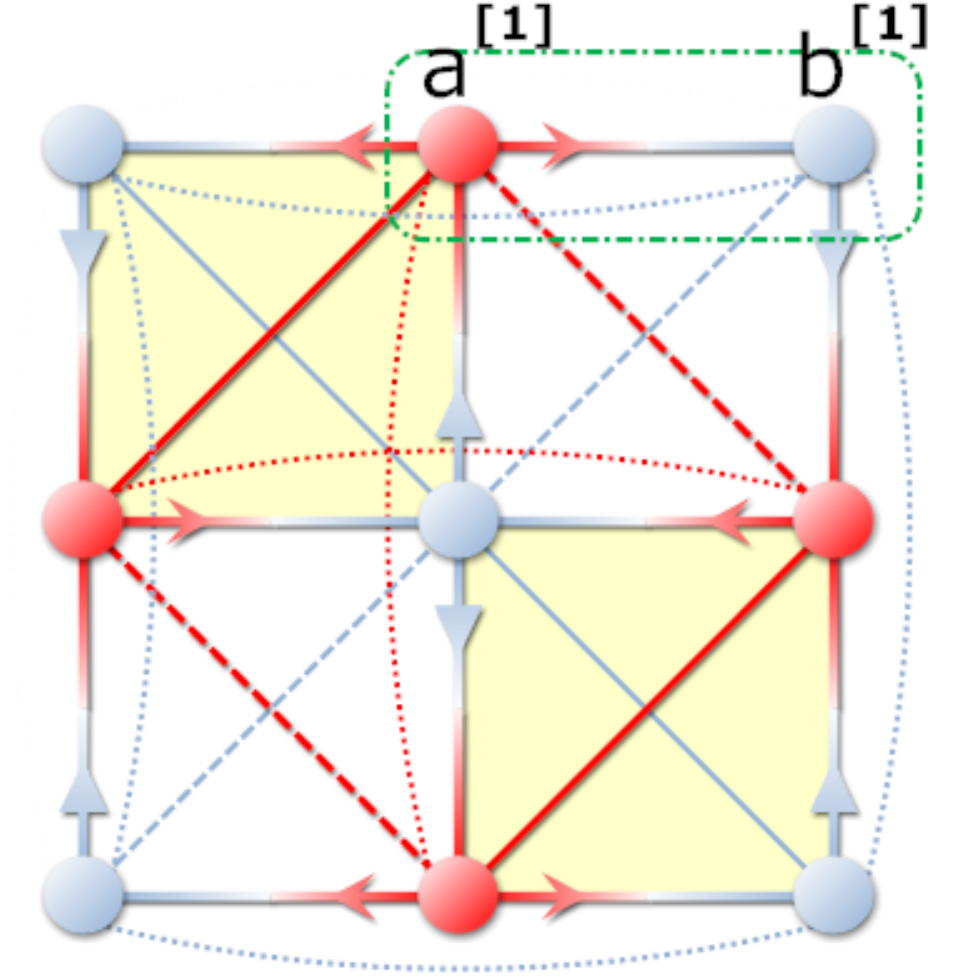}\label{fig:modelC2_layer1}}
\hspace*{0.4cm}
\subfigure[]{\includegraphics[width=0.20\columnwidth]{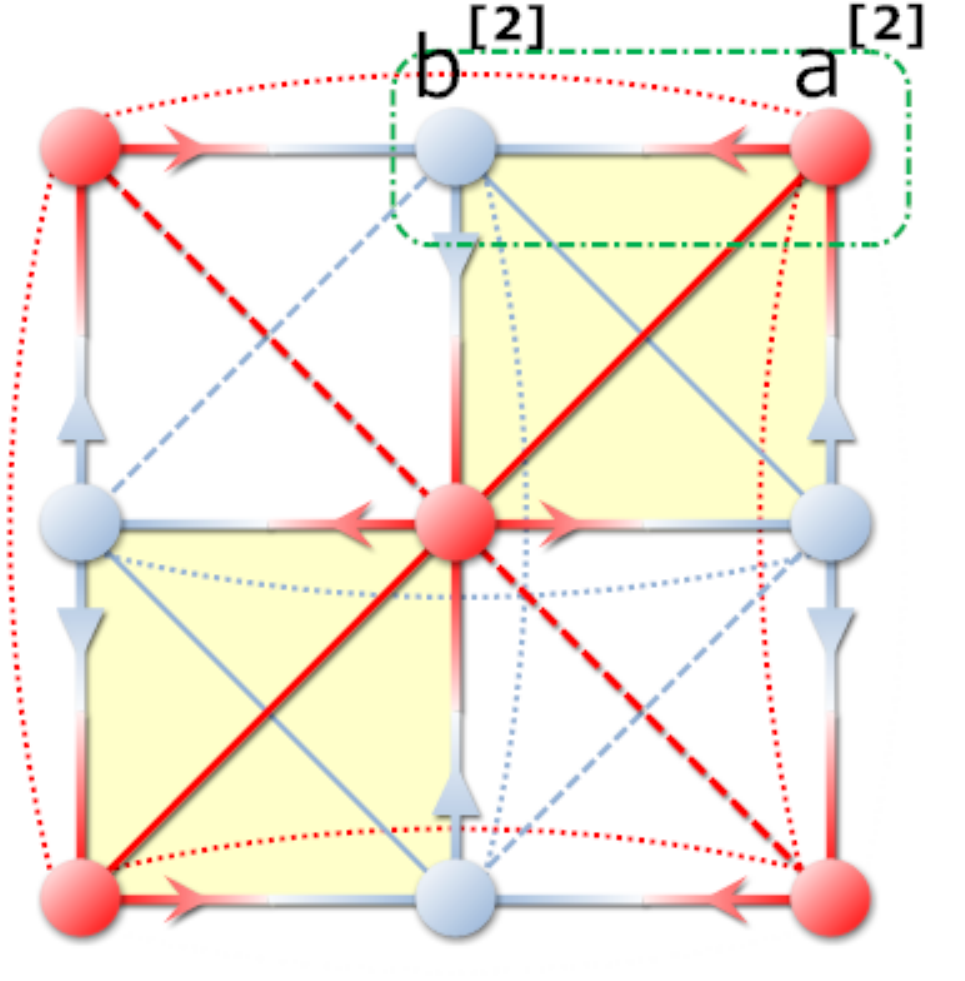}\label{fig:modelC2_layer2}}
\caption{(Color online) The two-orbital Chern number two model on a square lattice. (a) NN and NNNN hoppings. (b) NNN hoppings. (c) Band structure as a function of $k_{x}$ and $k_{y}$. (d) Edge states of this model. (e)-(f) Bi-layer checkerboard lattice models. Conventions are the same as in Figs. 1 and 2 of the main text.}
\end{center}
\end{figure}

As mentioned in the main text, the two-orbital Chern number $=2$ model on a square lattice can be separated into two layers of checkerboard models. For the bi-layer checkerboard model shown in Figs.~\ref{fig:modelC2_layer1} and \ref{fig:modelC2_layer2}, although in each layer the unit cell of a checkerboard lattice contains two sites, the unit cell of the bi-layer structure only has one lattice site (with two orbitals).
The Hamiltonian of each layer is
\begin{align}
H_{CB}^{[l]} &= t_{1}\sum_{\langle i,j \rangle } e^{i \phi_{ij}} a_{i}^{[l]\dagger} b_{j}^{[l]} +\sum_{\langle \langle i,j \rangle \rangle } t'_{ij} \left( a_{i}^{[l]\dagger}a_{j}^{[l]}+b_{i}^{[l]\dagger}b_{j}^{[l]} \right) +t_{3} \sum_{\langle \langle \langle i,j \rangle \rangle \rangle } \left( a_{i}^{[l]\dagger}a_{j}^{[l]}+b_{i}^{[l]\dagger}b_{j}^{[l]} \right) 
\label{HChern2a}
\end{align}
where $l=1,2$. By introducing the unitary transformation
\begin{align}
A_{i1}^{\dagger}=a^{[1]\dagger}, A_{i2}^{\dagger}=a^{[2]\dagger},
B_{i1}^{\dagger}=b^{[2]\dagger}, B_{i2}^{\dagger}=b^{[1]\dagger},
\label{transformationC2}
\end{align}
we recover the two-orbital Chern number $=2$ model on a square lattice.

\section{$k$-space Hamiltonian of the $N$-orbital $C=N$ model}

The $k$-space Hamiltonian of the $N$-orbital $C=N$ model reads
\begin{align}
H^{[C=N]}&= \sum_{\mathbf{k}}\sum_{l=1}^{N} \mathcal{F}_{\mathbf{k}}([2l-1]\phi)C_{\mathbf{k}}^{[l]\dagger}C_{\mathbf{k}}^{[l]} +\sum_{\mathbf{k}}\sum_{l=1}^{N} \left[ \mathcal{G}_{1\mathbf{k}}\left(2l\phi\right)C_{\mathbf{k}}^{[l+1]\dagger}C_{\mathbf{k}}^{[l]} + \mathcal{G}_{2\mathbf{k}}\left([2l+1]\phi\right)C_{\mathbf{k}}^{[l+2]\dagger}C_{\mathbf{k}}^{[l]}+\mathrm{H.c.} \right],
\label{HChernN}
\end{align}
where $C_{\mathbf{k}}^{[l]\dagger}$ is the single-particle creation operator of the $l$th orbital at momentum $\mathbf{k}$ with $1\le l \le N$ and we require
$C_{\mathbf{k}}^{[l]\dagger}$ satisfies periodic boundary conditions, i.e.,
$C_{\mathbf{k}}^{[N+1]\dagger}=C_{\mathbf{k}}^{[1]\dagger}$. The functions $\mathcal{F}$ and $\mathcal{G}$ are
\begin{align}
\mathcal{F}_{\mathbf{k}}(\Phi)&=2t_{2}\cos (k_{x}+k_{y}-\Phi), \notag \\
\mathcal{G}_{1 \mathbf{k}}(\Phi)&=t_{1}[\exp(ik_{x})+\exp(-ik_{y}+i \Phi)], \notag \\
\mathcal{G}_{2 \mathbf{k}}(\Phi)&=t_{2}\exp(ik_{x}-ik_{y}+i \Phi).
\end{align}

\end{widetext}
\end{appendix}


\begin{thebibliography}
\bibitem{chernnumber}
The Chern number of a many-body state is an integral invariant in the boundary phase space: $C={{1}\over{2\pi}}\int d\theta_x d\theta_y F(\theta_x,\theta_y)$, where $\theta_x$ and $\theta_y$ are the boundary phases in $x$ and $y$ directions, and the Berry curvature is given by $F(\theta_x,\theta_y)=\rm{Im}
\left(\left\langle {{\partial
\Psi}\over{\partial\theta_y}}\Big{|}{{\partial
\Psi}\over{\partial\theta_x}}\right\rangle -\left\langle {{\partial
\Psi}\over{\partial\theta_x}}\Big{|}{{\partial
\Psi}\over{\partial\theta_y}}\right\rangle\right)$.
\end{thebibliography}

\end{document}